\documentclass[journal]{IEEEtran}
\usepackage{amsmath,amsfonts}
\usepackage{array}
\usepackage[caption=false,font=normalsize,labelfont=sf,textfont=sf]{subfig}
\usepackage{textcomp}
\usepackage{stfloats}
\usepackage{url}
\usepackage{balance}
\usepackage{verbatim}
\usepackage{graphicx}
\usepackage{cite}
\usepackage{booktabs}
\usepackage{dcolumn}
\usepackage{algpseudocode}
\usepackage[ruled]{algorithm2e}
\usepackage{multirow}
\usepackage{enumerate}
\usepackage{enumitem}
\usepackage{float}
\usepackage{diagbox}
\usepackage{tcolorbox}
\usepackage{xcolor} % 引入xcolor包以支持更多颜色和自定义颜色
\newcolumntype{d}[1]{D{,}{,}{#1}}
\newcolumntype{.}[1]{D{.}{.}{#1}}
\newcommand{\tool}{\texttt{MoCo}}
\definecolor{DarkGoldenrod}{RGB}{106,90,205}
\definecolor{Ivory}{RGB}{248, 248, 255}
\begin{document}

\title{MoCo: Fuzzing Deep Learning Libraries\\via Assembling Code}

\author{Pin Ji, Yang Feng, Duo Wu, Lingyue Yan, Pengling Chen, Jia Liu, and Zhihong Zhao
        % <-this % stops a space
\thanks{P. Ji, Y. Feng, D. Wu, L. Yan, P. Chen, J. Liu, and Z. Zhao are with State Key Laboratory for Novel Software Technology, Nanjing University, Nanjing, China. Y. Feng is the corresponding author.}
\thanks{E-mail: pinji@smail.nju.edu.cn, fengyang@nju.edu.cn, \{duowu, lingyueyan, penglinchen\}@smail.nju.edu.cn, \{liujia, zhaozhih\}@nju.edu.cn}

% <-this % stops a space
}

% \markboth{IEEE Transactions on Software Engineering,~Vol.~14, No.~8, March~2024}%
% {Ji \MakeLowercase{\textit{et al.}}: MoCo: Fuzzing Deep Learning Libraries via Assembling Code}


\maketitle

\begin{abstract}
The rapidly developing deep learning (DL) techniques have been applied in software systems with various application scenarios.
However, they could also pose new safety threats with potentially serious consequences, especially in safety-critical domains.
While researchers are focusing on how to test DL models or domain-specific DL applications, only a little attention has been paid to testing DL libraries. 
DL libraries serve as the underlying foundation for DL systems, and bugs in them can have unpredictable impacts that directly affect the behaviors of DL systems. 
Previous research on fuzzing DL libraries still has limitations in the diversity of test inputs, the construction of test oracles, and the precision of detection. 
In this paper, we propose \tool, a novel fuzzing testing method for DL libraries via assembling code. 
The seed tests used by \tool~are code files that implement DL models, including constructing, training, and evaluating DL models in the most common real-world user scenarios. 
\tool~first disassembles the seed code file to obtain the template and code blocks, and then employs code block mutation operators (e.g., API replacement, random generation, and boundary checking) to generate more new code blocks adapted to the template.
By inserting context-appropriate code blocks into the template step by step, \tool~can generate a tree of code files with intergenerational relations. 
According to the derivation relations in this tree and the applied mutation operators, we construct the test oracle based on the execution state consistency. 
Since the granularity of code assembly and mutation is controlled rather than randomly divergent, we can quickly pinpoint the lines of code where the bugs are located and the corresponding triggering conditions. 
We conduct a comprehensive experiment to evaluate the efficiency and effectiveness of \tool~using three widely-used DL libraries~(i.e., TensorFlow, PyTorch, and Jittor).
During the experiment, \tool~detects 64 new bugs of four types in three DL libraries, where 51 bugs have been confirmed, and 13 bugs have been fixed by developers.
The experimental results demonstrate that \tool~is capable of generating high-quality tests and detecting different types of bugs, which helps developers improve the reliability of DL libraries.
\end{abstract}


\section{Introduction}
Nowadays, increasingly mature DL techniques have been applied to various types of software systems, and these DL systems have been used in safety and security critical application scenarios, such as autonomous driving~\cite{grigorescu2020survey}, intelligent security systems~\cite{hasan2018autonomous}, and surgical robots~\cite{esteva2019guide}. 
Typically, DL libraries are the basis for building DL systems, making it easier and more efficient to design, train, and deploy their core components—DL models. 
However, DL libraries are essentially software systems. 
They may suffer from software defects that result in serious consequences, such as models producing incorrect predictions, security vulnerabilities in the system, and leakage of private data. 

To date, some model-level and API-level DL library testing methods have been proposed. 
The model-level methods generate new DL models as test inputs, while the API-level methods generate new calling codes about a single or pair of APIs. 
Both LEMON~\cite{wang2020deep} and Muffin~\cite{gu2022muffin} represent model-level methods that detect bugs by comparing the outputs or the computational processes of generated DL models. 
They differ in that LEMON generates new DL models by adding or removing layers of the seed model and changing the value of the weight, and Muffin relies on a randomly generated abstract structure to select the appropriate layers for creating new DL models. 
For API-level methods, Wei et al. propose an API-level method, FreeFuzz~\cite{10.1145/3510003.3510041}, which extracts information from official documents and open-source code to identify common parameter settings and constraint rules. 
Deng et al. propose DeepREL, which infers relational APIs and compares their output values or execution states in order to identify inconsistencies~\cite{deng2022fuzzing}. 
The experimental results show that FreeFuzz and DeepREL perform well in testing TensorFlow and PyTorch.

Despite the promising progress, existing DL library fuzzing technologies still have the following limitations. 
First, the test input generation requires further improvement. 
The structures of the seed models or the abstract model structures greatly limit model-level methods. 
For example, LEMON can only add or remove existing layers of the seed model. 
Hence, the structure of the seed model limits the number of APIs tested and the number of generated models. 
Muffin restricts the available layer types for model construction to ensure the validity of the generated models.
For API-level methods, the granularity of fuzzing is either a single API or a pair of APIs.
The code they generated lacks the common context of the tested API and cannot closely resemble real application scenarios for DL libraries. 
Second, the reliability of the test oracle needs further exploration. 
Most current testing technologies perform cross-library or cross-API differential testing to address the test oracle problem. 
However, the implementation of difference testing presupposes that the DL libraries under test use the same front-end (e.g., Keras), and relational APIs have similar behaviors, resulting in limited libraries and APIs for adaptation. 
They also often share code logic (which may contain similar bugs) that make differential testing ineffective~\cite{10.1145/3510003.3510041}. 
Furthermore, only a few existing studies have reported the false positive rate of the proposed methods. 
Therefore, the time to verify the testing results and locate bugs is unpredictable.

To alleviate the limitations mentioned above, we propose \tool, a novel DL library fuzzing testing method via assembling code.
\tool~employs the code files for implementing DL models as seed tests, which are easier to modify and contain complete and rich semantics compared to both DL models and the calling code of API.
\tool~first disassembles the seed into the template and multiple code blocks and then applies a series of code block mutation operators (e.g., API replacement, boundary checking, and random generation) to generate more context-appropriate code blocks. 
After that, \tool~inserts them into the corresponding slots of the original code block in the template to ultimately produce a tree of code files for implementing various DL models. 
Based on the derivation relations in the tree and the mutation operators used, we define execution state consistency as the metamorphic relation for solving the test oracle problem.
The execution states consistency relation means that, guided by the derivation template, the execution state of the newly generated code file should be consistent with that of the inserted code block. 
Once a node in the code file tree violates this relation, we can quickly locate the line of code that may contain bugs and corresponding triggering conditions according to the derivation relation and the mutated portion.

To validate \tool, we experiment with three widely-used DL libraries, i.e., TensorFlow~\cite{tensorflow_github}, PyTorch~\cite{pytorch_github}
, and Jittor~\cite{jittor_github}.
We select 9 code files for building different DL models as seed tests, which cover RNN~(Recurrent Neural Network) and CNN~(Convolutional Neural Network) models with various parameter sizes.
In total, \tool~detects 64 new bugs in the new stable versions of three DL libraries.
To date, we have submitted these bugs to the developers of each DL library, 51 of which have been confirmed and 13 of which have been fixed by developers.
The experimental results demonstrate that \tool~has excellent testing efficiency and effectiveness.
Compared with existing model-level methods, \tool~can make full use of existing seed tests to cover more APIs and lines of code under test.
Compared with existing API-level methods, \tool~can report bugs with higher precision by constructing a more reliable test oracle. 
In addition, due to the feature of code assembly, \tool~allows developers to locate bugs and identify the triggering conditions quickly. 
% Moreover, \tool~can evaluate whether DL libraries behave reasonably in the face of abnormal parameter values, which makes up for the shortcomings of current methods.
% Compared with existing methods, \tool~can make full use of existing seed tests to cover more APIs and lines of code under test, and has higher precision in reporting bugs. 
% Therefore, \tool~can generate a large number of tests that contain more semantic and contextual information based on one seed, thus triggering more bugs that other fuzzing methods cannot detect. 
The main contributions of this paper can be summarized as follows:

\begin{enumerate}
    \item \textbf{Method.} We propose \tool, a novel DL library fuzzing testing method via assembling code. 
    \tool~can generate a tree of code files from one seed and detect bugs in DL libraries based on derivation relations. 
    % \tool~can fully tap the potential of seed tests to solve existing challenges. 
    
    \item \textbf{Tool.} We implement the proposed method in an open-source tool~\cite{moco-github}. 
    This tool can fully tap the potential of seed tests to solve existing challenges and assist developers in effectively and efficiently improving the reliability of various types of DL libraries.
    % This tool can disassemble seed tests, mutate code blocks, assemble code files, execute code files, and output error reports.
    \item \textbf{Study.} We conduct a comprehensive experiment to evaluate \tool. The experimental results demonstrate that \tool~detects 64 new bugs in the stable version of three DL libraries in total, 51 of which have been confirmed and 13 of which have been fixed by developers.
\end{enumerate}

\section{Background}
\subsection{Deep Learning Model}
A DL model is composed of many layers, each containing multiple neurons.
A neuron is the individual computing unit within the model, applying an activation function to its input and passing the result to other connected neurons. 
A DL model has at least three layers: an input layer, an output layer, and one or more hidden layers~\cite{pei2017deepxplore}. 
Each neuron in a layer has direct connections to neurons in the next layer. 
Each connection is bound to a weight parameter that represents the strength of the connection between neurons, learned from the training set given by the developers~\cite{wang2020deep}.
While the settings of the input and output layers are often related to the inputs and outputs of the selected dataset, the hidden layers are the focus of developers when designing their models.
Typically, developers design appropriate model structures based on actual requirements. 
For example, Convolutional Neural Networks (CNNs) extract features that focus more on local areas and have parameter-sharing characteristics that can greatly reduce computation costs, making them commonly used in image processing applications~\cite{li2021survey}. 
Recurrent Neural Networks (RNNs) have memory capabilities and are often used for sequential feature learning, such as speech recognition and machine translation~\cite{salehinejad2017recent}. 
Long-Short-Term Memory (LSTM) networks can retain important information from long sequences while ignoring irrelevant information, solving the gradient vanishing and exploding problems of RNNs and providing further performance improvements in various application scenarios~\cite{van2020review}.
The application of the DL model has greatly improved the advanced levels of speech recognition, computer vision, natural language processing, and other fields~\cite{lecun2015deep}. 

\subsection{Deep Learning Libraries}
DL libraries have emerged as indispensable tools in artificial intelligence, enabling researchers and practitioners to implement and explore complex neural network architectures efficiently. 
A DL library typically offers a wide range of predefined layers, activation functions~\cite{sharma2017activation}, various optimization algorithms~\cite{bottou2012stochastic, kingma2014adam}, and loss functions~\cite{wang2020comprehensive}, allowing developers to construct and customize DL models quickly. 
% These libraries also provide efficient implementations of various optimization algorithms, such as stochastic gradient descent~\cite{bottou2012stochastic} and Adam~\cite{}, which are essential for training DL models.
The research and industrial communities have developed and maintained numerous DL libraries. 
Well-known libraries include TensorFlow~\cite{abadi2016tensorflow}, PyTorch~\cite{paszke2019pytorch}, and Jittor~\cite{hu2020jittor}, which have gained widespread adoption and support due to their flexibility, scalability, and extensive community contributions.

\begin{figure*}[h]
  \centering
  \includegraphics[width=0.8\linewidth]{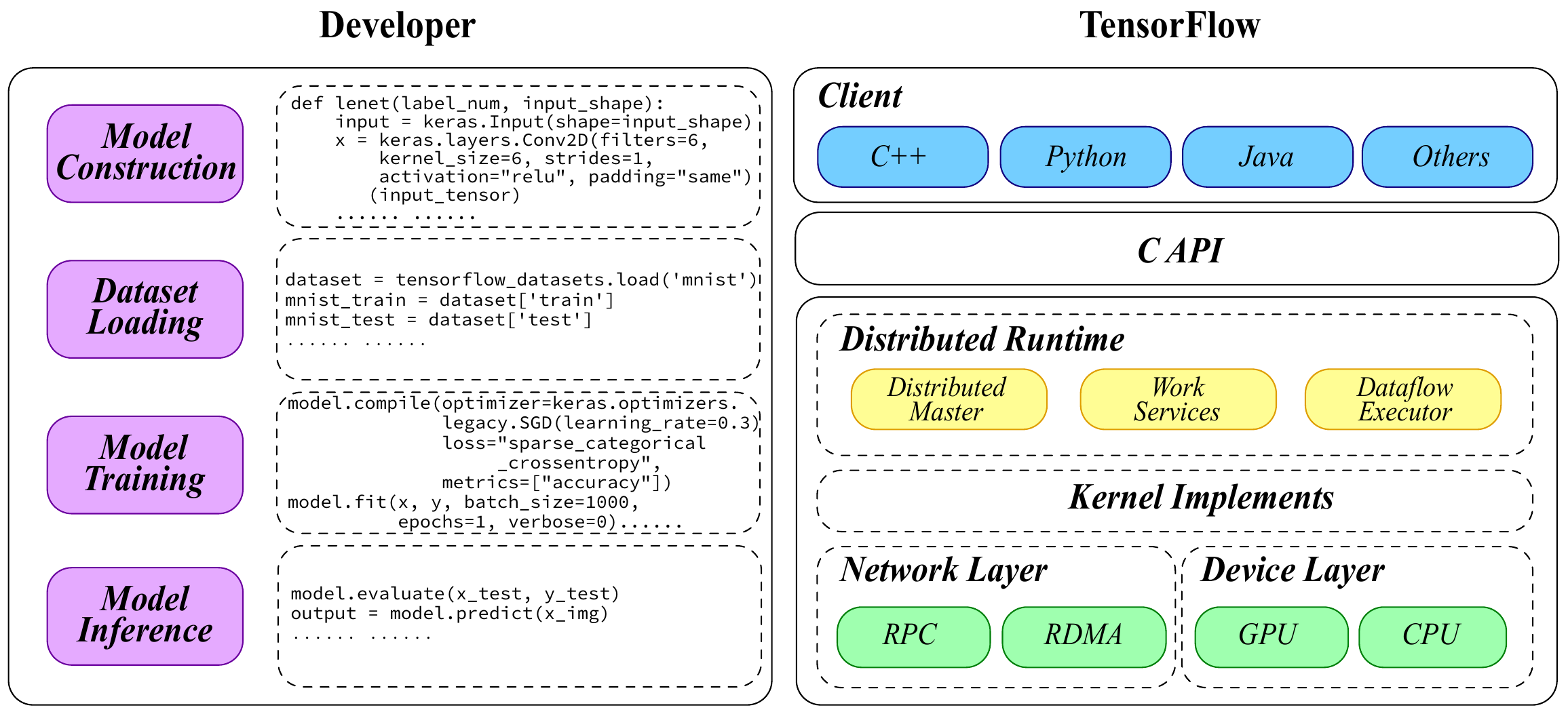}
  \caption{The architecture of DL library (TensorFlow)}
  \label{fig:DL-library}
\end{figure*}

We use TensorFlow as an example to illustrate the architecture of DL libraries in Figure~\ref{fig:DL-library}. 
The system architecture of TensorFlow is delineated by its C API, partitioning the entire system into two subsystems: the front end and the back end.
The front-end subsystem encompasses the programming model for constructing the computational graph. 
The back-end subsystem provides the running environment for executing the computational graph.
% Developers can utilize various APIs provided by DL libraries to implement a DL model, which includes model construction, dataset loading, model training, and model inference. 
The most common application scenario for developers to leverage the various APIs provided by DL libraries is to implement a DL model, which includes model construction, dataset loading, model training, and model inference. 
During the model construction, developers can easily specify the types and shapes of the input layer, hidden layers, and output layer by invoking the APIs. 
Therefore, the quality of code related to model implementation in DL libraries is crucial for the performance of DL systems.

\section{Approach}

\begin{figure*}[!t]
  \centering
  \includegraphics[width=\linewidth]{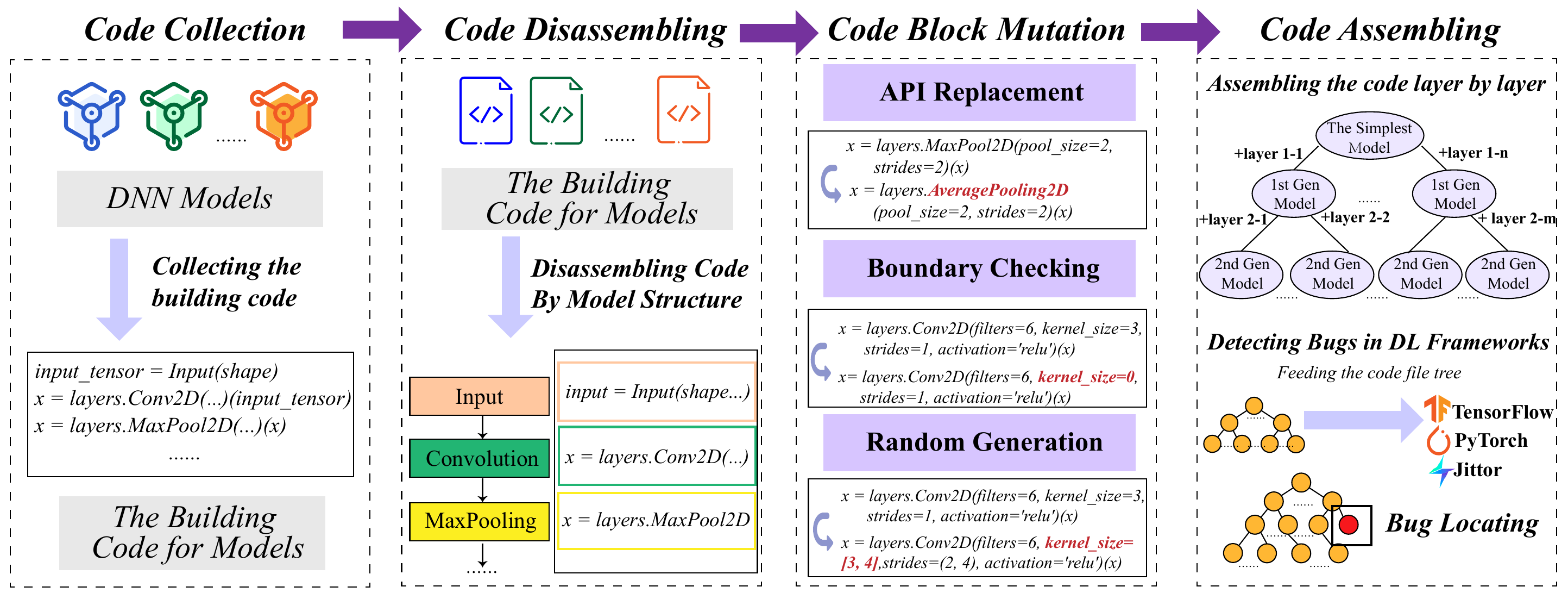}
  \caption{The Overview of \tool}
  \label{fig:overview}
\end{figure*}

% This section introduces the DL library fuzzing testing method via assembling code, which detect bugs by generating code in the most common application scenario of DL libraries. 
This section introduces the design and implementation of \tool, which aims to detect bugs in DL libraries by assembling code. 
% The code \tool~generates is tailored for common application scenarios, such as the construction, training, and evaluation of DL models based on fuzzy testing theory.
Figure~\ref{fig:overview} presents the overview of \tool. 
\tool~first disassembles the seed code file for implementing the DL model into the initial template and original code blocks. 
Then, \tool~employs three code block mutation operators (i.e., API replacement, random generation, and boundary checking) to generate more context-sensitive code blocks and generates the code derivation tree via assembling code. 
Finally, \tool~sequentially confirms the execution state of each node and identifies the code block where the bug is located according to the derivation relation and the mutation portion.

% \ji{I am not sure if the title of this subsection is appropriate}
\subsection{The Design Of \tool}

% \section{Implementation}
% This section introduces the fuzzing testing method via assembling code and contains the implementation details of this method. 
% \subsection{Overview}
Currently, there are two challenges in DL library testing: (1) lack of valid test inputs that cover more input space to trigger bugs and (2) lack of test oracles to reveal the bugs in DL libraries.
To address these challenges, we propose \tool, a novel fuzzing testing approach for DL libraries. 
The key idea of \tool~is to regard the seed code file as consisting of unchangeable skeleton code and multiple changeable code blocks related to specific requirements. 
The skeleton code is bare-bones but fully functional code containing all crucial elements, while the code blocks are used to enrich the semantics of the code file, and omitting them cannot affect the semantics and syntax correctness of the code file. 
Therefore, new code files can be generated by adding, deleting, and modifying the code blocks. 
Obviously, whether the newly generated code files can be executed properly is closely related to whether the inserted or modified code blocks are contextually and semantically syntactically correct. 
In this paper, we select the code files that implement the DL models as the seed tests.
Implementing the DL model is the most common application scenario of DL libraries, so the APIs involved are the most widely used and important. 
The potential bugs in these APIs can cause a huge negative impact on DL systems. 
% Therefore, we need to focus on testing the modules related to implementing DL models in the DL libraries.
To address the lack of valid test inputs, we design code disassembly and assembly methods and multiple suitable code mutation operators to improve the coverage of the input space. 
According to the derivation relation generated by mutating code blocks and assembling code, \tool~can easily solve the test oracle absence to achieve the detection and the location of bugs. 
% \tool~can make full use of the information from the seeds to complete the fuzzing and employ the characteristics of the assembling step to judge and locate the bug. 
% Due to the greater flexibility in modifying code compared to modifying deep learning models, \tool~can generate more diverse test inputs, covering more APIs and lines of code of DL libraries than model-level fuzzy testing methods. 
% Compared with the API-level methods that select a single API as the fuzzing seed, \tool~employs real data to complete the model training and evaluation, enabling testing more flow dependency patterns. 

\subsubsection{Code Disassembly}
\begin{figure}[htbp]
  \centering
  \includegraphics[width=\linewidth]{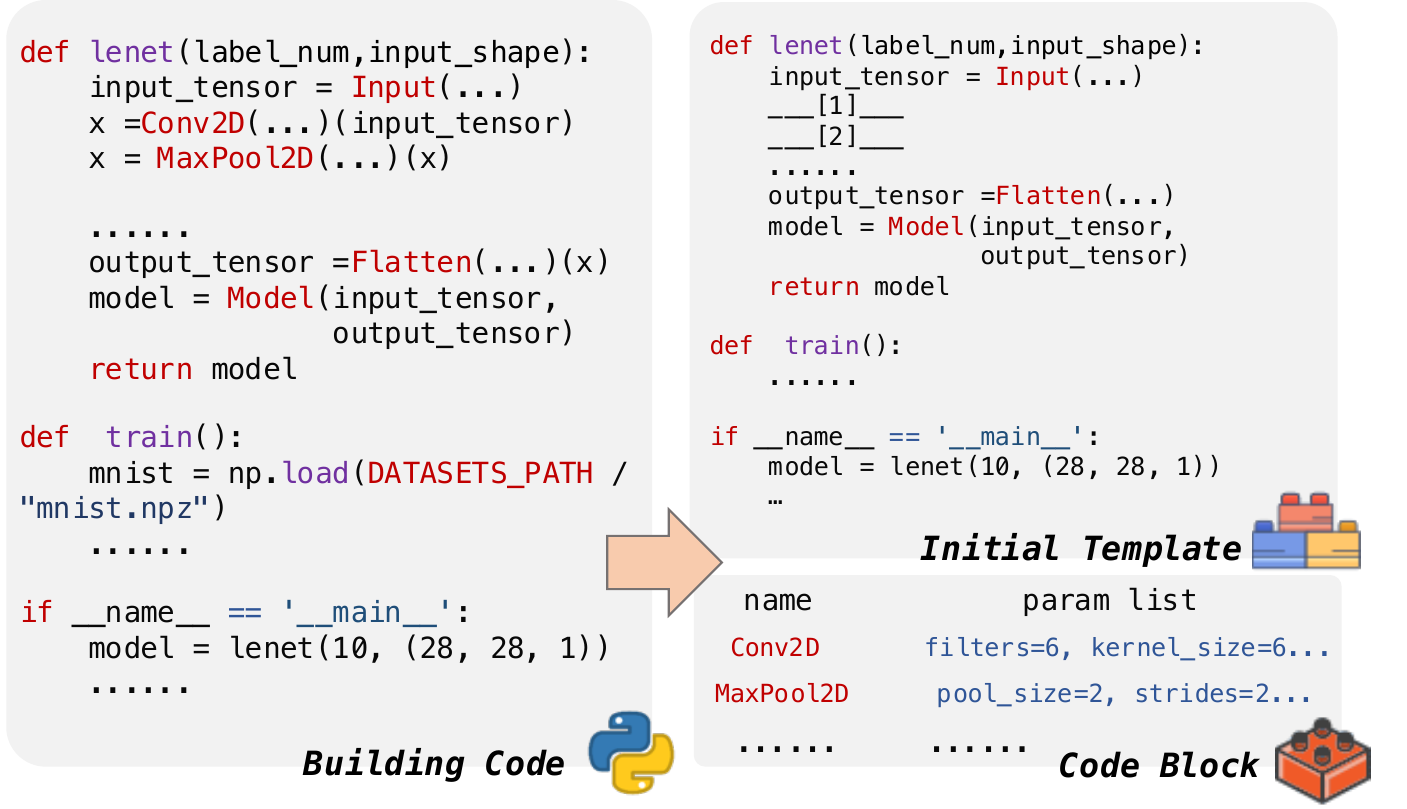}
  % \captionsetup{skip=0.2em}
  \caption{The process of code disassembly}
  \label{fig:disassemble}
\end{figure}

Based on the functional requirements for implementing DL models and the specific usages of DL libraries, we design a code disassembly method $\mathbf{D}$ for the code files that implement the DL models. 
The above code files consist of three parts: model construction, model training, and model evaluation. 
This method disassembles the runnable code file into an initial template $t_0$ that guides the generation of new code files and a set of code blocks $B_0$ that can be inserted into the template without affecting the semantic and syntactic correctness. 
Figure~\ref{fig:disassemble} shows the input and outputs of the code disassembly method. 

We define the initial template $t_0$ as the foundation for subsequent code assembly, consisting of the construction code corresponding to the input and output layers, slots where the code block can be inserted, and basic code for model training and validation. 
Therefore, the initial template for restricting the most basic context can be considered as derived from determining the suitable positions as slots in the skeleton code. 
Further, we define the code block $b$ as the assembly component, and the content is retrieved from the construction code corresponding to one hidden layer in the model construction. 
The code block retains rich semantic information about the original code, including API names and parameter values.
Together with the template, code blocks provide a basis and guidelines for subsequent code derivation to ensure the validity of the generated tests.
Given the seed code file $\mathbf{C}$, the code disassembly process can be formulated as:
$$
\mathbf{C} \stackrel{\mathbf{D}}{\longrightarrow} t_0, \mathbb{B}=\left\{b_1, b_2, \ldots, b_n\right\}
$$

\subsubsection{Code Derivation}
After disassembling code, we can obtain the initial template $t_0$, a set of original code blocks $\mathbb{B}$. 
% a group of mutated code block sets $\mathbf{M}(\mathbb{B}) = \left\{\mathbf{M}(b_1), \mathbf{M}(b_2), ... , \mathbf{M}(b_n)\right\}$. 
We define code assembly as the insertion of a code block into the corresponding slot of the initial template $t_0$ to generate a new code file. 
% During the step-by-step code assembling process, the template is also constantly updated, requiring constant refreshing of the context information of the next slot. 
To improve the test adequacy, we design three code block mutation operators: API replacement, boundary value check, and random generation.
These mutation operators can improve the diversity of the APIs and parameter value combinations covered by generated tests and check whether the DL libraries handle abnormal parameter values reasonably well. 
When code assembly is combined with code block mutation, the assembly path can be realized to evolve from a line to a tree diagram. 
% Given the initial template $t_0$ and a set of original code blocks $\mathbb{B}$, the output of the code assembling $\mathbf{A}$ is a code derivation tree $\mathbb{MT}$. 
Given the initial template $t_0$ and a set of original code blocks $\mathbb{B}$, we can obtain the derivation chain $\mathbb{C}_0 \stackrel{t_0+\mathbf{M}\left(b_1\right)}{\Rightarrow} \mathbb{C}_1 \stackrel{T_1+\mathbf{M}\left(b_2\right)}{\Rightarrow} \mathbb{C}_2 \ldots \mathbb{C}_{m-1} \stackrel{T_{m-1}+\mathbf{M}\left(b_m\right)}{\Rightarrow} \mathbb{C}_m$. 
We define the output of the code assembly $\mathbf{A}$ is a derivation tree $\mathbb{MT}$, and the whole process can be formulated as:
$$
\begin{aligned}
\mathbb{MT} & =\mathbf{A}\left(t_0, \mathbb{B}\right)=t_0+\mathbf{M}\left(b_1\right)+\ldots+\mathbf{M}\left(b_n\right) \\
& =\mathbb{C}_0 \cup \mathbb{C}_1 \cup \mathbb{C}_2 \ldots \mathbb{C}_{m-1} \cup \mathbb{C}_m
\end{aligned}
$$
Each node in the tree $\mathbb{MT}$ is a new code file to implement a DL model.
% The nodes are categorized into \textit{positive} and \textit{negative} based on the mutation operator used, indicating whether the corresponding code file should or should not work correctly.
During the derivation, \tool~generates new code files level by level and executes them sequentially at the end of each level generation. 
If a node at the $i_{th}$ level does not work correctly, it no longer has any children nodes. 
Therefore, the final number of tests generated by \tool~is related to the execution state of each node. 
% By comparing the execution of the code before and after inserting the code block, we can easily locate the API that may have bugs.
% By comparing the execution state of the nodes with the parent-child generation relation, \tool~can locate which inserted code block has the suspected bug.

\subsubsection{Test Oracle Construction}
Depending on the mutation operator used, we classify the nodes in the derivation tree $\mathbb{MT}$ into \textit{positive} type and \textit{negative} type, indicating whether the corresponding code file should or should not execute successfully without triggering any exceptions. 
Specifically, positive nodes are those where the semantics and syntax of the inserted code block are correct, such as those derived using random generation and API replacement; 
Negative nodes are those where the semantics and syntax of the inserted code block are incorrect, such as those where boundary value checking is used to generate parameter values that are out of the legal range of values. 
Since the premise of node derivation is that its parent node can execute normally, and the derivation process is controlled by the template, which ensures that the insertion position of the code block is contextually appropriate, the expected execution state of each node should be consistent with the execution state of the inserted code block. 
We formalize this execution state consistency relation as the metamorphic relation to address the test oracle problem:
$$
    \mathbf{S}(b) \stackrel{\mathbf{S}(t)=\textit{true}}{\Rightarrow} \mathbf{S}(\mathbf{A}(t,b))
$$
where $\mathbf{S}$ denotes the execution state of the code, $t$ denotes the template used for this code derivation, and $b$ denotes the code block inserted into the template $t$. 
We set $\mathbf{S}(c)$ to \textit{true} if the code $c$ can be executed normally and set $\mathbf{S}(c)$ to \textit{false} if not. 
By predicting the execution state of each node, we can easily detect bugs in the DL libraries. 
Moreover, we can quickly locate the line of code where the bug is located and clarify the trigger conditions according to the derivation relation between nodes with parent-child relations and the mutation portion in the inserted code block. 

\subsection{Template Generation}
To ensure the effectiveness of the code disassembly method $\mathbf{D}$, we first collect the code files using representative DL libraries (e.g., TensorFlow~\cite{tensorflow_github}, PyTorch~\cite{pytorch_github}, and Jittor~\cite{jittor_github}) to implement various DL models for analysis. 
After investigation, we find two common approaches for implementing DL models: (1) serially calling functional APIs and (2) customing classes to inherit from parent classes in DL libraries (e.g., torch.nn.Module).
For these two programming implementation modes, \tool~adopts distinct treatment methods. 
The first implementation approach combines the definition of the model structure and the computation process into one, and the order of API calls corresponds to the model structure, so the code of the first approach can be directly disassembled by line in order. 
The second approach divides the initialization definitions of the layers and the building of the model structure into two functions, so \tool~first completes the mapping between the code defining the layer and the code calling the corresponding variables before disassembling the code. 
To ensure that the generated templates and code blocks are syntactically compliant and correct, \tool~inserts the seed code blocks into the slots of the initial template in order and verifies that the generated code file works after each assembly step.

\subsection{Code Block Mutation Operators}
To improve the diversity of tests generated by \tool, we design the following three mutation operators to mutate code blocks for fuzzing DL libraries thoroughly: 
\begin{enumerate}[leftmargin=*]
    \item API Replacement (AR): AR belongs to the API mutation, which means replacing the original API with a new API that can be used in the same context. We define API functional similarity to mean that after replacing API $A_1$ with API $A_2$ in a code file, the code file will still work properly. 
    This definition indicates that $A_1$ and $A_2$ have similar semantics and can implement similar functions. This operator can effectively increase the number of APIs covered by generated tests.
    
    \item Random Generation (RG): RG belongs to the parameter mutation, meaning a parameter value in the API is randomly changed. The new value is generated at random within the legal range. This operator can increase the probability of triggering bugs by covering more combinations of parameters. Compared to the other methods that assign random values for each parameter in the API, this operator only changes one parameter value, making it easier to pinpoint what triggered the bug. 
    
    \item Boundary Checking (BC): BC belongs to the parameter mutation, which means replacing the original parameter value with an illegal value near the legal value boundary. This operator can check whether the DL library has detected and processed abnormal parameter values. In DL, handling illegal values in the inputs of API is crucial to ensure the stability, robustness, and security of the DL models. Therefore, it is necessary to use this operator in DL library testing.
    
\end{enumerate}

\begin{tcolorbox}[title = {Prompts for parsing the official documents}, fonttitle = \sffamily, fontupper = \sffamily, fontlower = \itshape, width=0.95\linewidth, colback=Ivory, colframe=DarkGoldenrod]
  \footnotesize
  \textbf{Round 1}
  \renewcommand{\labelitemi}{-}
  \renewcommand{\labelitemii}{·}
  \begin{itemize}[leftmargin=0.5cm]
      \item You are an expert in the field of deep learning. Next, I'll give you information about the parameters of the \textit{[API name]} from the official documentation of the deep learning library \textit{[DL Library name]}. Please read the text I give you next and answer the questions I ask you. Once you understand the requirements, please answer only 'Yes.' Then, I will continuously ask you several questions.  
      \item \textit{[information]}
  \end{itemize}

  \textbf{Round 2}
  \renewcommand{\labelitemi}{-}
  \renewcommand{\labelitemii}{·}
  \begin{itemize}[leftmargin=0.5cm]
      \item Please give me the dtype and the range for each parameter in the following format. The type indicates the data type of each parameter, such as int, float, boolean, and string. The range indicates the legal range of values for each parameter. The specific ranges are represented in Python code. If not mentioned in the text, it is marked as ``No mention.'' \\
      \{
      \begin{itemize}[label={}]
        \item Parameter name:
        \item dtype:
        \item range:
        \end{itemize}
        \}
  \end{itemize}

  \textbf{Round 3}
  \renewcommand{\labelitemi}{-}
  \renewcommand{\labelitemii}{·}
  \begin{itemize}[leftmargin=0.5cm]
      \item Please provide the structure and shape for each parameter in the following format. The structure indicates the type of data structure that stores a collection of values for the input parameter, such as integer, list, tuple, and n-dimensional array (i.e., tensor). Shape indicates the shape or number of dimensions of the parameter, and it is related to the number of integers contained in the tuple or the list. It is marked as ``No mention'' if not mentioned in the text. \\
        \{
      \begin{itemize}[label={}]
        % \item Parameter NO:
        \item Parameter name:
        \item structure:
        \item shape:
       \end{itemize}
        \}
  \end{itemize}

  \textbf{Round 4}
  \renewcommand{\labelitemi}{-}
  \renewcommand{\labelitemii}{·}
  \begin{itemize}[leftmargin=0.5cm]
      \item Please provide the constraints between the parameters in the following format. Parameter 1 and parameter 2 refer to the two parameters for which a constraint relationship exists. The specific constraints are represented in Python code. The Constrain NO is auto-incremental. \\
    \{
      \begin{itemize}[label={}]
        \item Constrain NO:
        \item Parameter 1:
        \item Parameter 2:
        \item Constrain:
        \end{itemize}
        \}
  \end{itemize}

\end{tcolorbox}

Implementing three mutation operators depends on the definitions of APIs, the parameter lists of APIs, and the definitions and the legal value ranges of API parameters. 
We employ the crawler tool \textit{Beautiful Soup}~\cite{bs-crawler} to obtain the detailed descriptions of the APIs from the official documentation, which includes the definition of the API, the initialisation code of the API and the parameter information. 
However, the required information is not directly available from official documents, especially the data type, value range, structure and shape of the parameters, and the possible constraints between them. 
Therefore, we employ the large language model \textit{GPT-4} to parse the crawled documents, and the prompts used in this process are shown in the text box above. 
After many attempts, we find that multiple rounds of interrogation with GPT performed significantly better than extracting all the information at once.
Therefore, we go through four rounds to obtain the data type, range, shape, structure, and constraints of each parameter. 
We require \textit{GPT-4} to express the constraints between parameters and the legal value ranges in Python, which is more beneficial to the accurate implementations of mutation operators.
% We employ crawlers and Natural Language Processing (NLP) techniques to obtain API-related information from the official documentation and process them into a structured, easy-to-read storage form~(e.g., YAML and JSON). 

The RG and BC operators only require the parameter list and the corresponding value ranges, but the AR operator also requires functional similarity between APIs.
We design a measurement method $\tau$ to calculate the functional similarity.
$\tau$ measures the similarity between two APIs in two dimensions: the definitions of APIs and the parameter lists of APIs.
First, $\tau$ employs the transformer~\cite{han2021transformer} model to convert the definitions of API $A_1$ and API $A_2$ into sentence embeddings~\cite{li2018word} $V_1$ and $V_2$, and then calculates the cosine similarity of $V_1$ and $V_2$ according to the following formula~\cite{lahitani2016cosine}:

\begin{footnotesize}
$$
CosSim(V_1, V_2) = \frac{V_1 \cdot V_2}{\|V_1\|\|V_2\|}=\frac{\sum_{i=1}^n V_{1_i} \times V_{2_i}}{\sqrt{\sum_{i=1}^n\left(V_{1_i}\right)^2} \times \sqrt{\sum_{i=1}^n\left(V_{2_i}\right)^2}}
$$ 
\end{footnotesize}

The similarity $Sim_{def}$ between the definitions of API $A_1$ and API $A_2$ is equal to the cosine similarity of $V_1$ and $V_2$. 
Then, $\tau$ treats each parameter in the parameter lists $P_1$ and $P_2$ as separate words and selects the Word Correctness Rate ($WCR$) to measure the similarity $Sim_{para}$ between the parameter lists of API $A_1$ and API $A_2$.
% $WCR$ is often used to evaluate speech recognition results, taking into account both the content and the location of the words. 
The calculation of $Sim_{para}$ is related to the \textit{Levenshtein} distance, which indicates the minimum number of substitutions, deletions, or insertions required to change a sequence into another sequence~\cite{yujian2007normalized}.
As shown in the following formulation, we define $Sim_{para}$ as equal to the percentage of parameters in the hypothesis parameter list $P_2$ whose position and data type match the reference parameter list $P_1$:

\begin{footnotesize}
$$
WCR(P_1, P_2)=\frac{(N_1-D-S)}{N_1}=\frac{C}{(S+D+C)}
$$
\end{footnotesize}
where $N_1$ is the number of parameters in $P_1$, $D$ is the number of deletions, $S$ is the number of substitutions, and $C$ is the number of identical parameters. 
The above procedure for calculating the functional similarity between API $A_1$ and API $A_2$ can be formulated as:

\begin{footnotesize}
$$
Sim_{A_1A_2} = \frac{Sim_{def} + Sim_{para}}{2} = \frac{CosSin(V_1, V_2) + WCR(P_1, P_2)}{2}
$$
\end{footnotesize}

After the above steps, as shown in the text box below, we can obtain the list of functional similarities, the list of parameter information, and the list of constraint information for each API.
Based on the information obtained, the AR operator needs to determine which API to replace the original API. 
We use the roulette wheel selection~\cite{shukla2015comparative} based on probabilistic selection in the field of genetic algorithms as the API selection strategy.
The roulette wheel method is a random selection method used in optimization algorithms, which emulates the process of natural selection.
In this method, individuals with higher fitness are more likely to be selected, similar to the survival of the fittest principle in biological evolution.
In this study, the perimeter of the roulette wheel is set as the sum of all similarity values, and each API occupies the interval corresponding to its similarity value in random order. 
Compared with simply using the similarity threshold to select APIs, this method can exhibit a high probability of selecting APIs with higher similarity to ensure the validity of generated tests and cover more APIs to improve the diversity of generated tests. 
After that, the AR operator randomly generates a number between zero and the sum of similarity and selects the API as the target API in whichever interval it falls.
During the code derivation, we randomly select one of the AR and RG operators to mutate the original code block, after which we perform the BC operator based on the newly generated code block to produce multiple code blocks required for negative nodes. 
We set the parameter $times_{mt}$ to determine the number of times the AR and RG operators can mutate one code block.

\begin{center}
\begin{tcolorbox}[title = {tf.keras.layers.LSTM.yaml}, fonttitle = \sffamily, fontupper = \sffamily, fontlower = \itshape, width=0.95\linewidth, colback=Ivory, colframe=DarkGoldenrod]
  \footnotesize
\textbf{Similarity:}
\renewcommand{\labelitemi}{-}
  \renewcommand{\labelitemii}{·}
  \begin{itemize}[leftmargin=0.5cm]
  \item tf.keras.layers.LSTMCell: 0.7150709480047226
  \item  tf.keras.layers.SimpleRNN: 0.644281268119812
  \item tf.keras.layers.GRU: 0.6300555363945339
  \item ......
  \end{itemize}
\textbf{Parameters:}
\renewcommand{\labelitemi}{-}
  \renewcommand{\labelitemii}{·}
\begin{itemize}[leftmargin=0.5cm]
\item activation: 
\begin{itemize} [label={}]
    \item default: tanh
    \item dtype: tf.string
    \item enum:
    \begin{itemize}[label={-}]
        \item tanh
        \item None
    \end{itemize}
\end{itemize}
    
\item activity\_regularizer:
\begin{itemize} [label={}]
    \item default: None
    \item dtype: tf.string
\end{itemize}
\item ......
\end{itemize}

\textbf{Constrains:}
\begin{itemize}[leftmargin=0.5cm]
    \item Parameter 1: unit\_forget\_bias,
    \item Parameter 2: bias\_initializer,
    \item Constrain: \\  if unit\_forget\_bias is True: \\ bias\_initializer = 'zeros'
    \item ......
\end{itemize}
\end{tcolorbox}
\end{center}

\subsection{Tree Pruning And Bug Detection}
\begin{figure}[h]
  \centering
  \includegraphics[width=\linewidth]{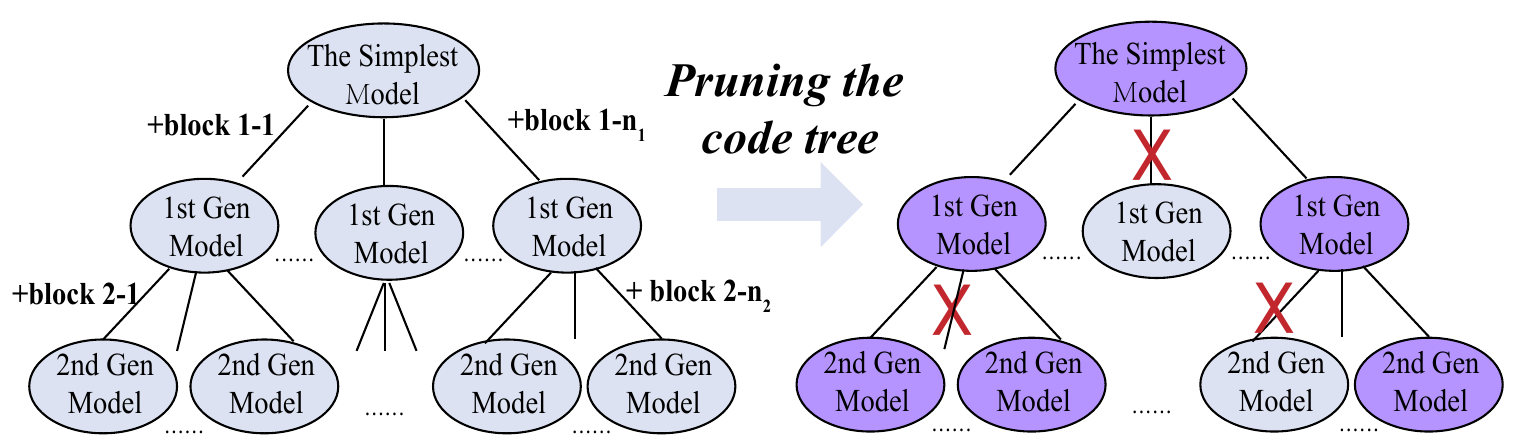}
  \caption{pruning the code derivation tree}
  \label{fig:code-assembling}
\end{figure}

From Figure~\ref{fig:code-assembling}, we infer that each increase in the number of code block mutations can produce an exponential increase in the number of nodes in the derivation tree, which can also result in a high time cost for testing. 
Therefore, we design a code tree pruning strategy based on code diversity to improve testing efficiency while ensuring test adequacy.
This strategy is inspired by beam search, an algorithm to optimize the outputs and constrain the search space~\cite{djukanovic2019beam}. 
When the nodes in each level finish executing, we categorize the nodes that have run properly into equivalence classes and then retain 50\% of each class for further derivation. 
Since negative nodes should not have child nodes, this pruning strategy is designed for the subsequent derivation of positive nodes. 
In this strategy, we measure the code diversity from two aspects: API diversity and parameter diversity. 
Therefore, we combine the mutation operators to classify the equivalence classes from the names of APIs and the values of parameters. 
Specifically, for the API mutation operator, we group nodes based on the replacement API in the inserted code block. If the same API is used to replace the original API, then they are grouped into the same equivalence class.
% Specifically, for the API mutation strategy, we group that nodes that represent inserted code blocks that are replaced with the same API into the same equivalence class. 
% For the parameter mutation, we define the equivalence class as the legal value, minimum value, maximum value and illegal value of the mutated parameter. 
% For bool type variants, divide equivalence classes into true and False.
For the parameter mutation operators, we divide the different classification rules according to the types of parameters to be mutated.
For example, equivalence classes of numeric types can be divided into mutated values that belong to the legal range, the illegal range, or equal to the boundary value;
equivalence classes of \textit{boolean} types are divided into \textit{TRUE} and \textit{FALSE}; equivalence classes of enumeration types are divided into all options. 
If we do not adopt the pruning strategy, the tree generated by \tool~is similar to an n-ary tree.
The number of positive nodes in this tree is approximately equal to $1 + n + n^2 + ... + n^{(m-1)} + n^m$, where $n$ is the mutation times of the code block, and $m$ is the number of original code blocks. 

\section{Evaluation}
We implement the above approach upon Python and experiment with three commonly used DL libraries~(i.e., TensorFlow, PyTorch, and Jittor), 9 seed code files for building DL models, and 4 datasets to evaluate \tool. 
Furthermore, we conduct a bug study to analyze the characteristics of \tool~and the bugs in the tested DL libraries. 
We have made \tool~and all raw data open source~\cite{moco-github}.
All experiments are performed on a Ubuntu 18.04, one 12-core CPU processor with 2.40 GHz, and one NVIDIA GeForce RTX 3090 GPU. 
In this study, we address the following research questions:
\begin{itemize}[leftmargin=*]
    \item \textbf{Efficiency.} In a specific time, how many tests can \tool~generate? How does \tool~perform in terms of covered APIs and code lines?
    \item \textbf{Effectiveness.} How many bugs are detected by \tool? How precise is \tool~in reporting bugs? Where do false positives come from?
    \item \textbf{Bug Study}. What types of bugs does \tool~detect in commonly used DL libraries?

\end{itemize}

\subsection{Deep Learning Libraries}
In this paper, we select three widely used DL libraries, i.e., TensorFlow, PyTorch, and Jittor, as our subjects under test. 
TensorFlow is developed by Google Brain's Machine Intelligence team for studying machine learning and neural networks~\cite{tensorflow_github}. 
The main feature of TensorFlow is the use of a data flow diagram to complete the numerical calculation~\cite{abadi2016tensorflow}.
PyTorch is developed by Facebook~\cite{pytorch_github}, and it can offer two advanced functionalities: (1) powerful tensor computations with GPU acceleration and (2) deep neural networks built on a tape-based automatic differentiation system~\cite{paszke2019pytorch}. 
Jittor is developed by a research team at Tsinghua University. 
The distinctive feature of Jittor lies in its utilization of just-in-time compilation and meta-operators, enabling the generation of customized high-performance code for models~\cite{hu2020jittor}.
As a new DL library, it has great applications and development potential~\cite{jittor_github}.
Table~\ref{tab:DLL-info} shows more statistics information about these three DL libraries.
The column \textbf{\#SLOC} presents the number of source code lines, and \textbf{\#Star} presents the number of stars in their official GitHub repositories.

\begin{table}[htbp]
\caption{Statistics information of DL libraries under test}
\label{tab:DLL-info}
\centering
\begin{tabular}{cccccc}
\toprule
\textbf{DL Library}&\textbf{Version}&\textbf{\#SLOC}&\textbf{\#Star}&\textbf{Language}\\
\hline
TensorFlow&2.12.0&954,440&176k&C/C++, Python\\
PyTorch&2.0.0&707,635&68.2k&C/C++, Python\\
Jittor&1.3.7.16&77,978&2.8k&C/C++, Python\\
\bottomrule
\end{tabular}
\end{table}

\subsection{Models And Datasets}
In our experiments, we use two methods to collect the code for implementing DL models: (1) collecting the code files from GitHub and (2) collecting the DL models
and then using MMdnn~\cite{mmdnn_github} to transform them.
In the end, we select 9 code files as the seed tests to test three DL libraries.
The models to be implemented by these code files have been widely used in much existing research~\cite{li2019boosting, pham2019cradle, wang2020deep}. 
These nine models enhance the diversity of the seed data by covering different structures of DL models (e.g., CNN model and RNN model), the scales of models (e.g., small and large model in terms of the number of layers and weights), and various types of input sets (e.g., images, and sequential data). 
Table~\ref{tab:seed-info} shows the statistics information of the DL models corresponding to the code files.

\begin{table}[htbp]
\caption{Statistics information of seed tests}
\label{tab:seed-info}
\footnotesize
\centering
\begin{tabular}{c|c|c|c|c}
\toprule
\textbf{Model}&\textbf{Dataset}&\textbf{\#Weights}&\textbf{\#Layers}&\textbf{Type}\\
\hline
LeNet&MNIST&75k&10&CNN\\
\hline
SqueezeNet&ImageNet&744k&14&CNN\\
\hline
PointNet&Stock-Price&2,899k&15&CNN\\
\hline
AlexNet &CIFAR-10&20,345k&17&CNN\\
\hline
VGG19&CIFAR-10&143,667k&26&CNN\\
\hline
MobileNet&CIFAR-10&4,221k&32&CNN\\
\hline
% DenseNet&ImageNet&5,364k&121&CNN\\
% \hline
% VGG16&CIFAR-10&138,358k&23&CNN\\
% \hline
% InceptionV3&ImageNet&685,321k&159&CNN\\
% \hline
ResNet18&ImageNet&11,685k&71&CNN\\
\hline
% ResNet50&ImageNet&123,835k&159&CNN\\
% \hline
GoogleNet&ImageNet&6,999k&153&CNN\\
\hline
% Xception&ImageNet&22,913k&126&CNN\\
% \hline
LSTM&Stock-Price&3k&5&RNN\\
% \hline
% BiLSTM&Stock-Price&8k&303&RNN\\
% \hline
% GRU&Stock-Price&13k&13&RNN\\
\bottomrule
\end{tabular}
\end{table}

\subsection{Baseline Selection}
We select two types of DL library testing methods as baselines: a model-level method, Muffin~\cite{gu2022muffin}, and an API-level method, DeepREL~\cite{deng2022fuzzing}. 
Muffin can generate diverse DL models as test inputs and relies on data analysis in the training phase for differential testing to detect bugs, representing the advanced level of model-level fuzzing methods. 
Muffin employs a top-down generation algorithm that first generates the abstract model structure and then generates each layer based on predefined supported APIs. 
Muffin also supports testing with seed models, but it just loads the seed model without using any mutation operators. 
Muffin can be used to test DL libraries with Keras as a front-end, such as TensorFlow~\cite{tensorflow_github}, Theano~\cite{Theano-github}, and CNTK~\cite{CNTK-github}. 
DeepREL infers potential API relations based on their syntactic/semantic information and then performs fuzzing on the verified relational APIs to find inconsistencies. 
The potential API relations are the value equivalence (given the same inputs, two APIs can output the same result) and the state equivalence (given the same input, the running states of two APIs are the same). 
The experimental results have demonstrated that DeepREL has better testing performance than FreeFuzz~\cite{10.1145/3510003.3510041} on previous versions of TensorFlow and PyTorch. 

Since Muffin is a cross-library differential testing tool, running it on at least two DL libraries is essential to complete the bug detection.
However, the version of TensorFlow we chose to test is 2.12.0, and its adapted Keras does not have a corresponding version for Theano~\cite{Theano-github} and CNTK~\cite{CNTK-github}. 
% Thus, we only compare the diversity of generated tests with Muffin, and do not compare the number of detected bugs. 
Compared to DeepREL, \tool~is closer to Muffin in terms of generating tests, both of which aim to cover DL models with more diverse structures. 
Therefore, we compare the test case generation ability between Muffin and \tool~to evaluate the testing efficiency of \tool, and analyze the bug detection ability of Muffin based on the information provided by its repository. 
% Therefore, comparing the diversity of generated tests between Muffin and \tool~is more appropriate to evaluate the testing efficiency of \tool.
From the perspective of fuzzing, the goal of DeepREL is the same as the goal of~\tool, which is to cover as much input space as possible to trigger bugs. 
Therefore, we select DeepREL as the baseline to compare the effectiveness of detecting bugs based on seeds with \tool. 

\subsection{Evaluation Metrics}
To evaluate \tool, we use the following metrics:

\textbf{Number of Covered APIs:} There are a large number of APIs in existing DL libraries, so it is important to show the number of covered APIs as a measure of test adequacy, especially whether those commonly used APIs in the application level of DL libraries can be covered. We use $\#Cov_{API}$ to represent this metric. 

% It is used to measure the efficiency of \tool.\\
\textbf{Number of Covered Code Lines:} The number of covered code lines is a widely used test adequacy metric for testing traditional software systems. It refers to the total number of covered code lines in the C/C++ source code of each DL library during the execution of tests. 
We use $\#Cov_{line}$ to represent this metric.
$\#Cov_{line}$ and $\#Cov_{API}$ are used to evaluate the efficiency of \tool. 
We use LCOV~\cite{lcov-github} and Bazel~\cite{bazel} to calculate the $\#Cov_{line}$ when testing TensorFlow and use the tool provided in the repository of PyTorch when testing PyTorch~\cite{torch-coverage}. 

\textbf{Number of Detected Bugs and Precision:} We use the number of bug-related reports $\mathbb{R}_{bug}$, the number of detected bugs \#$Bug$, and the precision of \tool~to evaluate the effectiveness of \tool. 
We calculate the precision of \tool~employing the following formula: 
% \textbf{Precision: }
\begin{equation}
    \text {P}_{\text{bug}}=\frac{\sum_{r \in R}\{\operatorname{Check}(r)\}}{|R|}
\end{equation}
where $R$ is the set of suspected bug reports output by the testing tool, and $|R|$ is the size of $R$. 
We set $Check(r)$ to true if the suspected bug report $r$ is related to the bug of the DL library and set $Check(r)$ to false if it is not. 
% The precision of~\tool~can be regarded as the percentage of suspected bug reports that are really related to the bug of the DL library. 
In previous studies, the precision of DL library testing methods has received little attention. 
% However, it is actually a very important evaluation metric, and higher precision enables developers to consume less time to find real bugs.

\section{Results Analysis And Discussion}
\subsection{The Efficiency Of \tool}
Based on the seed code files and datasets described above, we employ \tool~to perform fuzzing testing on three DL libraries. 
Firstly, we set the number of mutation times for one code block $times_{mt}$ to three, four and five to conduct the test case generation three times, and compare the types and numbers of equivalence classes in the generated code file trees. 
% Firstly, we conduct three rounds of test case generation by setting the number of mutation times for one code block $times_{mt}$ as three, four and five, and compare the types and numbers of equivalence classes in the generated code file trees. 
Considering the testing adequacy and efficiency, we select four for the complete experiment.
To reduce the effect of randomness on the experimental results, we run \tool~five times with the same configuration and take the rounded results of the average values as the final experimental results. 
% In experiments, we set the number of mutation times for one code block $times_{mt}$ to four. 
Table~\ref{tab:test-gen} shows the number of new tests for each DL library generated by~\tool, denoted as \#$Test$, and the average running time cost for each test generation and execution denoted as $Avg_{t}$. 
% To highlight the testing efficiency of \tool, we compare \tool~with LEMON.
Table~\ref{tab:cov}~shows the number of covered APIs and the number of covered code lines in the C/C++ source code of each DL Library during the execution of the tests generated by \tool~and Muffin. 
Since we focus on using DL Libraries to construct DL models, the covered APIs we count in Table~\ref{tab:cov} are those in the modules that are most relevant to neural networks, i.e., \textit{tf.keras.layer}, \textit{torch.nn}, and \textit{jittor.nn}, which have 169 APIs, 144 APIs, and 72 APIs, respectively. 

% \begin{table}[htbp]
% \caption{Test generation summary for each seed}
% \scriptsize
% \centering
% \label{tab:test-gen}
% \begin{tabular}{c|c|c|c|c|c|c}
% \toprule
% \multirow{2}{*}{\textbf{Model}}&\multicolumn{2}{c|}{\textbf{TensorFlow}}&\multicolumn{2}{c|}{\textbf{PyTorch}}
% &\multicolumn{2}{c}{\textbf{Jittor}}\\
% \cline{2-7}
% &\textbf{\#$Test$}&\textbf{$Avg_{t}$}&\textbf{\#$Test$}&\textbf{$Avg_{t}$}&\textbf{\#$Test$}&\textbf{$Avg_{t}$}\\
% \hline
% LeNet&264,478&0.25&28,963&0.03&19,829&0.60\\
% \hline
% SqueezeNet&765,625&0.12&14,648,167&0.03&4,849,378&0.01\\
% \hline
% PointNet&236,095&0.44&9,407&0.13&42,126&0.93\\
% \hline
% AlexNet&258,195&0.31&178,308&0.24&412,117&0.11\\
% \hline
% VGG19&452,493&0.30&1,513,315&0.06&325,128&0.04\\
% \hline
% MobileNet&206,682&0.28&2,083,370&0.02&4,808,293&0.01\\
% \hline
% ResNet18&335,835&0.47&122,490&0.57&45,915&0.71\\
% \hline
% GoogleNet&132,058&0.18&60,872&0.07&116,314&0.16\\
% \hline
% LSTM&2,270&2.90&1,005&0.08&1,820&0.24\\
% \bottomrule
% \end{tabular}
% \end{table}

\begin{table}[htbp]
\caption{Test generation summary for each seed}
\scriptsize
\centering
\label{tab:test-gen}
\begin{tabular}{c|c|c|c|c|c|c}
\toprule
\multirow{2}{*}{\textbf{Model}}&\multicolumn{2}{c|}{\textbf{TensorFlow}}&\multicolumn{2}{c|}{\textbf{PyTorch}}
&\multicolumn{2}{c}{\textbf{Jittor}}\\
\cline{2-7}
&\textbf{\#$Test$}&\textbf{$Avg_{t}$}&\textbf{\#$Test$}&\textbf{$Avg_{t}$}&\textbf{\#$Test$}&\textbf{$Avg_{t}$}\\
\hline
LeNet&27,690&0.25&16,147&0.03&20,250&0.60\\
% 0.024 29,524
\hline
SqueezeNet&48,864&0.12&2,616,341&0.03&1,967,305&0.01\\
% 0.0031 2,391,484
\hline
PointNet&10,549&0.44&4,613&0.13&51,910&0.93\\
% 0.0019
\hline
AlexNet&33,286&0.31&164,068&0.24&313,682&0.11\\
%0.000677
\hline
VGG19&300,663&0.30&972,442&0.06&205,378&0.04\\
% 0.0000068
\hline
MobileNet&108,619&0.28&1,761,359&0.02&3,548,863&0.01\\
% 0.000000318
\hline
ResNet18&87,596&0.47&122,490&0.57&45,915&0.71\\
% 0.0403 9,841
\hline
GoogleNet&29,676&0.18&77,208&0.07&38,866&0.16\\
\hline
LSTM&1,233&2.90&632&0.08&1,665&0.24\\
% 0.31 121
\bottomrule
\end{tabular}
\end{table}

As seen from Table~\ref{tab:test-gen}, \tool~can generate a tremendous number of code files for implementing new DL models based on a seed code file. 
\tool~generates an average of 72,020 new tests on TensorFlow, 637,256 new tests on PyTorch, and 688,204 new tests on Jittor based on each seed.
\tool~generates more tests on PyTorch and Jittor than Tensorflow because PyTorch and Jittor require more lines of code when constructing the same model, leading to an increase in the number of original code blocks that can cause the derivation tree to have more levels and nodes.
% An increase in the number of original code blocks means that the derivation tree generated by \tool~can have more levels and nodes. 
% which leads to an exponential increase in the number of nodes. 
The running time of the tests mainly includes the time for code generation, model construction, and training the model. 
Since we do not require the precision of the model, \tool~only randomly selects 100 data from the training set and sets the epoch to once during the training phase to control the running time of the entire testing process.
The running time of a test is related to the structural complexity of the constructed model, the number of weights included, the training data, and the execution state of each node.
Therefore, there is a difference between the running time of tests generated based on seed code files that implement different types and sizes of DL models.
The characteristics of DL libraries can also make a difference in the running time of tests.
For example, PyTorch is a DL library based on dynamic graphs and dynamically allocates and frees memory as needed, which makes it faster than TensorFlow.

\begin{table}[htbp]
\caption{Test coverage for each DL library}
\footnotesize
\centering
\label{tab:cov}
\begin{tabular}{c|c|c|c|c}
\toprule
\multirow{2}{*}{\textbf{Tool}}&\multicolumn{2}{c|}{\textbf{TensorFlow}}&\multicolumn{2}{c}
{\textbf{PyTorch}}\\
\cline{2-5}
&\#$Cov_{API}$&\#$Cov_{line}$&\#$Cov_{API}$&\#$Cov_{line}$\\
\hline
seed&12&15,540&13&43,554\\
\hline
\tool&\textbf{95}&\textbf{19,016}&\textbf{72}&\textbf{50,533}\\
\hline
Muffin-ran&59&18,053&\textbackslash&\textbackslash\\
\bottomrule
\end{tabular}
\end{table}

As shown in Table~\ref{tab:cov}, the code block mutation operators implemented in \tool~can effectively increase the number of covered APIs. 
The code coverage of tests generated by \tool~does not increase as significantly as the API coverage because the API and the primitive operator are in a one-to-many relation, i.e., different APIs can employ the same primitive operator to perform the computation. 
Muffin, which uses a random generation algorithm, selects an appropriate API from a predefined API pool, so the number of APIs covered is limited by the number and diversity of APIs in this pool. 
By reading the source code, we find that this predefined API pool contains 59 APIs.
Thus, the data in Table~\ref{tab:cov} shows the test generation ability of Muffin in the ideal situation. 
We reproduce the experiments of Muffin on four datasets (e.g., MNIST, ImageNet, Stock-Price, and CIFAR-10) using the docker image provided by its GitHub repository. 
When the number of generated models is set to 50, the total running time of Muffin is 292 minutes, and the average execution time for one testing round is 87.6 seconds, significantly slower than \tool.
% and its total runtime for generating 50 models is xxx, with an average execution and comparison time of xxx for one model. 
The other model-level fuzzing methods, such as LEMON, generate new models by removing and adding layers already in the seed models and require that the input shape and the output shape of the selected layer be the same. 
Therefore, the number of APIs covered by the generated tests is closely related to the number of APIs covered by the seed models.
Compared with them, \tool~employs multiple mutation operators and the code assembly method to enrich the semantics of generated tests, which are designed to effectively address the limitations of the existing test input generation methods and improve testing efficiency.

\subsection{The Effectiveness Of \tool}
We employ the number of bugs detected and the precision to clearly demonstrate the effectiveness of \tool~in detecting bugs.
Table~\ref{tab:bug-report} shows the number of bug-related reports output by \tool~and the precision when using different seed code files. 
We use DeepREL to test all the APIs in the modules most commonly used to build neural networks, i.e., \textit{tf.keras.layer} and \textit{torch.nn}, which have 169 APIs and 144 APIs, respectively. 
The APIs tested by \tool~are also in the above modules. 
Since DeepREL does not support Jittor, we do not compare the performance of DeepREL with \tool~when testing Jittor.
The key configuration parameter of DeepREL is the number of times each API pair is tested.
We use the optimal configuration provided by its authors to conduct the first round experiment.
In the second round experiment, we increase the number of tested times to 30,000 for TensorFlow and 60,000 for PyTorch, and control its total running time to be the same as the total running time of \tool. 
% Then we increase the number of times to 30,000 for TensorFlow and 60,000 for PyTorch and perform fuzzing testing seven times to keep its running time consistent with \tool~in the second round experiment. 

\begin{table}[htbp]
\caption{Reported error summary for each seed}
\footnotesize
\centering
\label{tab:bug-report}
\begin{tabular}{c|c|c|c|c|c|c}
\toprule
\multirow{2}{*}{\textbf{Model}}&\multicolumn{2}{c|}{\textbf{TensorFlow}}&\multicolumn{2}{c|}{\textbf{PyTorch}}
&\multicolumn{2}{c}{\textbf{Jittor}}\\
\cline{2-7}
&$\mathbb{R}_{bug}$&$P_{bug}$&$\mathbb{R}_{bug}$&$P_{bug}$&$\mathbb{R}_{bug}$&$P_{bug}$\\
\hline
LeNet&8&57.1\%&5&71.4\%&3&75.0\%\\
\hline
SqueezeNet&3&60.0\%&7&63.6\%&7&77.8\%\\
\hline
PointNet&5&62.5\%&5&71.4\%&6&75.0\%\\
\hline
AlexNet&13&76.4\%&6&54.5\%&6&85.7\%\\
\hline
VGG19&5&55.6\%&6&66.7\%&4&80.0\%\\
\hline
MobileNet&6&60.0\%&5&71.4\%&4&80.0\%\\
\hline
ResNet18&4&66.7\%&6&54.5\%&7&77.8\%\\
\hline
GoogleNet&7&58.3\%&5&62.5\%&7&70.0\%\\
\hline
LSTM&14&77.8\%&2&66.7\%&3&75.0\%\\
\hline
\textbf{Avg}&7.2&65.6\%&5.2&63.5\%&5.2&77.0\%\\
\bottomrule
\end{tabular}
\end{table}

We deduplicate the outputs of both \tool~and DeepREL, merging reports with the same API and similar exception information. 
In the first round experiment, DeepREL outputs 18 bug reports about TensorFlow, 4 of which are related to bugs, and outputs 71 bug reports about PyTorch, 1 of which are related to bugs. 
On the same testing time budget with \tool, DeepREL outputs 35 more reports about TensorFlow than in the first round, but no new bugs are detected, and outputs 10 more reports about PyTorch, which contain 5 new bugs. 
Both \tool~and DeepREL detect two bugs about \textit{tf.keras.layers.LayerNormalization}, and the remaining bugs detected by DeepREL do not overlap with the bugs detected by \tool. 
The experimental results in Table~\ref{tab:bug-report} and Table~\ref{tab:bug-type} show that \tool~outperforms DeepREL in terms of the number of bugs detected, the types of bugs detected, and the precision. 
The performance difference between \tool~and DeepREL is discussed further in Section~\ref{sec:bug-study}. 
For the effectiveness of Muffin, we analyse the outputs of Muffin when testing TensorFlow 2.0.0 with four datasets by using the scripts provided in its GitHub repository. 
The results show that Muffin only detects one bug of TensorFlow about \textit{BinaryCrossentropy}. 
Despite using the same parameter settings, other bugs mentioned in its paper have not been detected. 
Also, we find that models generated by Muffin suffer from incompatibility with TensorFlow 2.0.0 and later versions. 
For example, \textit{tf.placeholder} has been removed in TensorFlow 2.0.0 and later versions.
% Since its repository and paper do not provide detailed information about the bugs detected by Muffin, we identify the bugs by searching through its generated log files using "tensorflow" as a keyword.

As shown in Table~\ref{tab:fp}, we further analyze the source and the corresponding number of false positives in the error reports output by \tool.
% False positives are mainly from situations where the code fails to run properly and is not caused by the bug of the DL Library. 
Specifically, this situation is mainly caused by the following four reasons: (1) wrong training configuration, (2) syntax errors in the code, (3) semantic errors in the code, and (4) exhausted resources such as RAM or video memory.
Of these, syntax errors and semantic errors account for the largest proportion, reaching 77.4\%. 
This is because \tool~generates parameter values of APIs according to the official documentation, which can ensure that they are syntactically correct when running alone but does not guarantee that they can always fit the context when inserted into the template. 

\begin{table}[htbp]
\caption{Number of different types of false positives}
\footnotesize
\centering
\label{tab:fp}
\begin{tabular}{c|c|c|c|c}
\toprule
\textbf{DL}&\textbf{Training}&\textbf{Synatx}&\textbf{Semantic}&\textbf{Lack of}\\
\textbf{Library}&\textbf{Configuration}&\textbf{Error}&\textbf{Error}&\textbf{Resources}\\
\hline
TensorFlow&11(16.9\%)&26(40.0\%)&20(30.8\%)&8(12.3\%)\\
\hline
PyTorch&0(0.0\%)&7(25.9\%)&16(59.3\%)&4 14.8\%)\\
\hline
Jittor&0(0.0\%)&4(28.6\%)&9(64.3\%)&1(7.1\%)\\
\hline
Total&11(10.4\%)&37(34.9\%)&45(42.5\%)&13(12.3\%)\\
\bottomrule
\end{tabular}
\end{table}

\subsection{Bug Study}
\label{sec:bug-study}
% In that subsection, we select three representative bugs of tested DL Libraries for further analysis.
% We further deduplicate and categorize the bug-related reports output by \tool~and DeepREL. 
Table~\ref{tab:bug-type} shows the bug type and corresponding quantities detected by \tool~and DeepREL. 
We categorize the bugs detected by \tool~and DeepREL into the following four types: (1) If the code implementation is inconsistent with the official documentation, we define it as an inconsistency bug and denote the number of bugs of this type as \textbf{\#ICBug}.
(2) If the DL library has unreasonable allocation and reclamation of resources such as graphics memory, we define it as a performance bug and denote the number of bugs of this type as \textbf{\#PerBug}.
(3) If the DL library does not throw an exception to an illegal parameter value of the API and the code still continues to run or even crash, we define it as a boundary check bug and denote the number of bugs of this type as \textbf{\#BouBug}. 
(4) If the source code of the DL library has implementation errors (such as logic errors, API misuse, etc.), we define it as an implementation bug and denote the number of bugs of this type as \textbf{\#ImpBug}. 
The bugs detected by \tool~have been submitted to the developers of each DL library, and these issues can be viewed on our project website~\cite{moco-github}. 
We sent bug reports directly to the developers of Jittor through email and have received their responses to confirm the bugs.
Next, we select some representative bugs and analyze them in detail.

\begin{table*}[htbp]
\caption{The number of various types of bugs detected by \tool~and DeepREL}
\label{tab:bug-type}
\footnotesize
\centering
\begin{tabular}{c|c|c|c|c|c|c|c|c|c|c}
\toprule
\textbf{Library}&\multicolumn{2}{c|}{\textbf{\#ICBug}}&\multicolumn{2}{c|}{\textbf{\#PerBug}}&\multicolumn{2}{c|}{\textbf{\#BouBug}}&\multicolumn{2}{c|}{\textbf{\#ImpBug}}&\multicolumn{2}{c}{\textbf{\#Total}}\\
\hline
\multirow{2}{*}{Jittor}&MoCo&DeepREL&MoCo&DeepREL&MoCo&DeepREL&MoCo&DeepREL&MoCo&DeepREL\\
\cline{2-11}
&1&\diagbox{}{}&1&\diagbox{}{}&27&\diagbox{}{}&10&\diagbox{}{}&39&\diagbox{}{}\\
\hline
\multirow{2}{*}{TensorFlow}&MoCo&DeepREL&MoCo&DeepREL&MoCo&DeepREL&MoCo&DeepREL&MoCo&DeepREL\\
\cline{2-11}
&2&2&1&0&7&2&1&0&11&4\\
\hline
\multirow{2}{*}{PyTorch}&MoCo&DeepREL&MoCo&DeepREL&MoCo&DeepREL&MoCo&DeepREL&MoCo&DeepREL\\
\cline{2-11}
&6&1&0&0&6&4&2&1&14&6\\
\bottomrule
\end{tabular}
\end{table*}

\begin{figure}[h]
  \centering
  \includegraphics[width=\linewidth]{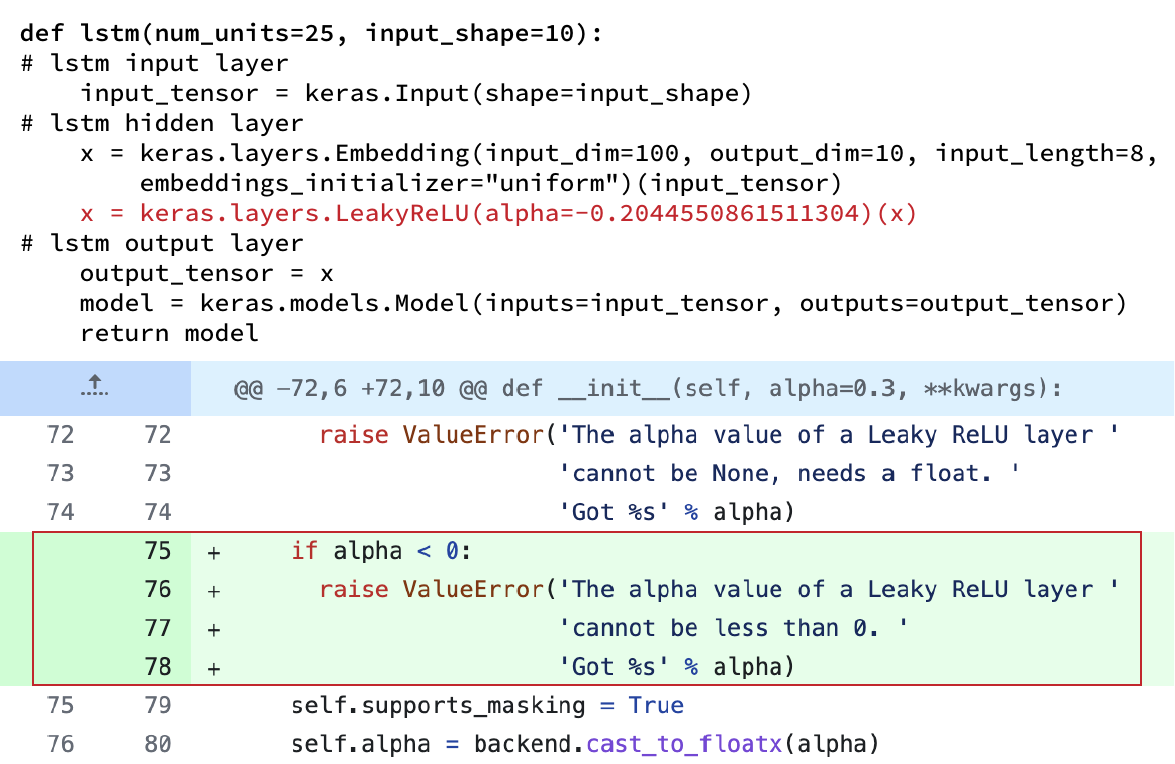}
  % \captionsetup{skip=0.3em}
  \caption{Boundary check bug in tf.keras.layers.LeakyReLU}
  \label{fig:bcbug}
  % \vspace{-1.0em}
\end{figure}

\textbf{Boundary Checking Bug.}
As shown from the code snippet using TensorFlow in Figure~\ref{fig:bcbug}, \tool~mutates \textit{x= keras.layers.Dropout(rate=-0.1)} in the seed to \textit{x = keras.layers.LeakyReLU(alpha=-0.2044)}.
By running the generated code, \tool~finds that even if the value of \textit{alpha} is less than 0, the code still completes the model construction, which is contrary to the legal value range of this parameter.
TensorFlow should perform boundary value checks for parameters of \textit{LeakyReLU} and throw the correct exception in time when the value is illegal.
After we submitted this bug, it has been fixed by the developers. 
% Like inconsistency bugs, FreeFuzz is more concerned with legal parameter values of APIs, making it difficult to trigger such bugs. 

\begin{figure}[h]
  \centering
  \includegraphics[width=\linewidth]{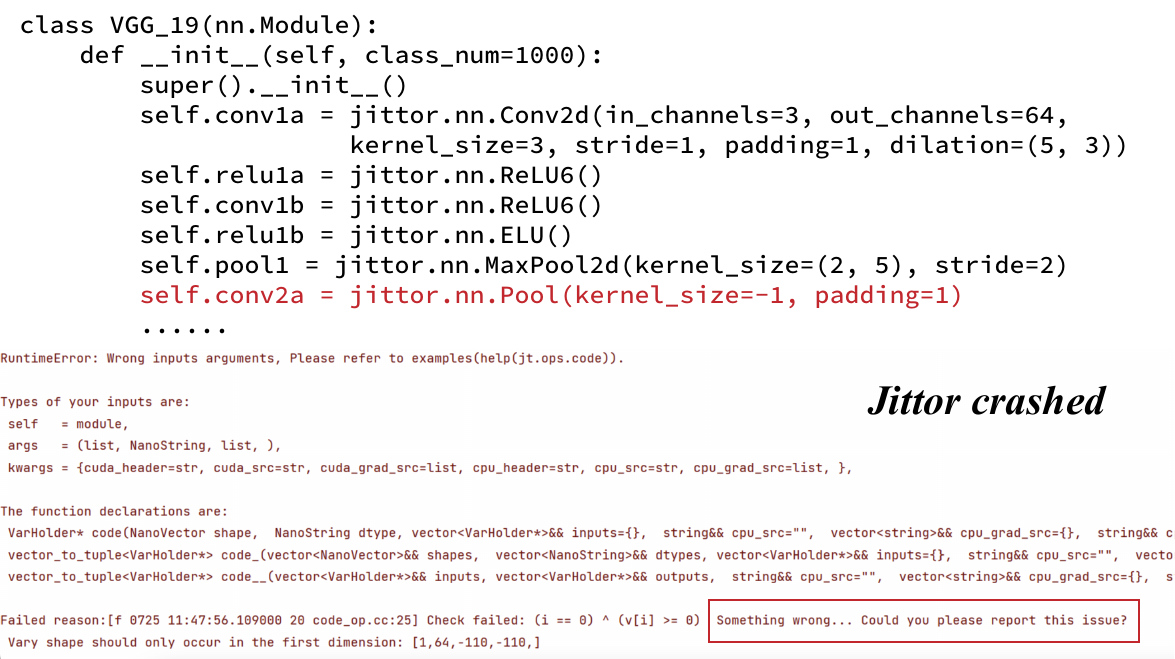}
  % \captionsetup{skip=0.3em}
  \caption{Boundary check bug in jittor.nn.Pool}
  \label{fig:bcbug-jt}
\end{figure}

As shown from the code snippet in Figure~\ref{fig:bcbug-jt}, the same type of bug presented in \textit{jittor.nn.Pool} causes serious consequences. 
Jittor does not perform boundary value checks for the value of \textit{kernel\_size}, resulting in the primitive computation operator crashing. 
% DeepREL cannot detect bug in \textit{tf.keras.layers.LeakyRelu}, we analyze this because it has a high probability of using the parameter values collected from documentation and open source repositories that work fine, which are difficult to trigger such bugs. 
The number of boundary bugs detected by DeepREL is significantly lower than \tool~because DeepREL has a high probability of randomly generating parameter values from a defined range of legal values, making it difficult to trigger this type of bug. 
% DeepREL can only detect bugs of this type if one of the relational APIs performs a value check on the corresponding parameter.
DeepREL can only detect bugs of this type if a parameter takes an illegal value and the API used for comparison performs a value check on this parameter.

% \vspace{-1.0em}

\begin{figure}[h]
  \centering
  \includegraphics[width=\linewidth]{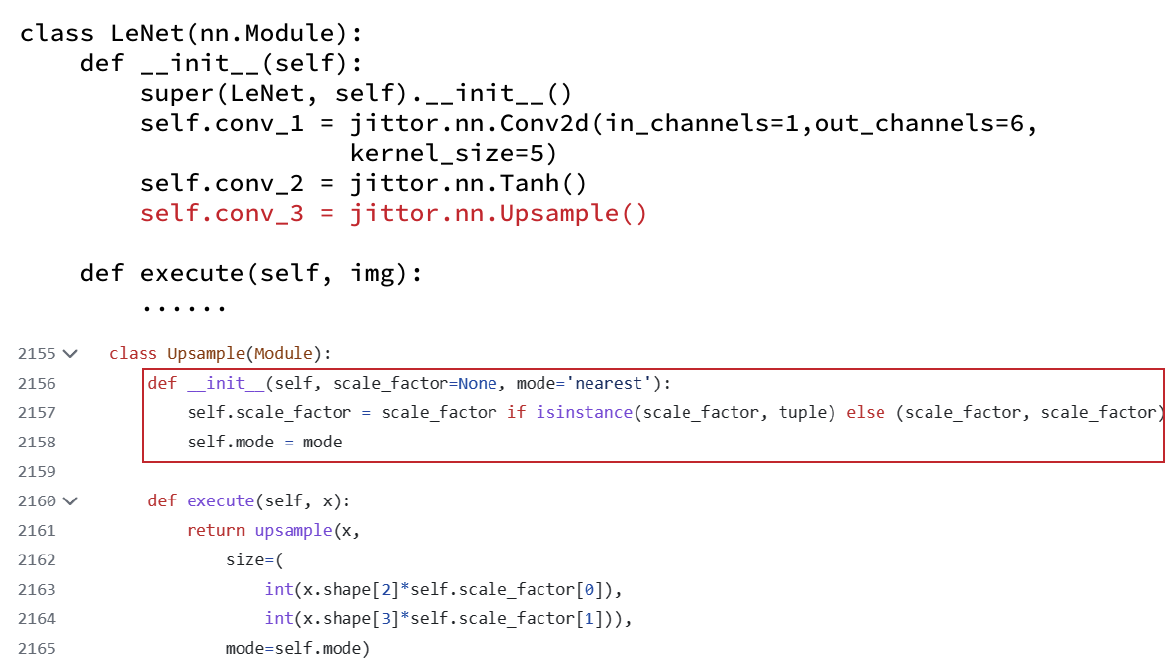}
  % \captionsetup{skip=0.3em}
  \caption{Implementation bug in jittor.nn.Upsample}
  \label{fig:impbug-jt}
\end{figure}

\textbf{Implementation Bug.} As shown from the code snippet using Jittor in Figure~\ref{fig:impbug-jt}, \tool~mutates \textit{jittor.nn.MaxPool2d()} in the seed to \textit{jittor.nn.Upsample()}. 
The code fails to run properly and throws a TypeError, indicating ``an unsupported operation type(s) for int and NoneType.''
% This bug occurs because Jittor does not check the value of \textit{scale\_factor} and even uses the default value \textit{None} for subsequent operations. 
This bug occurs because Jittor does not set a legal value as the default value of \textit{scale\_factor}, does not determine whether the value of \textit{scale\_factor} is empty, or even uses \textit{None} for subsequent operations. 
PyTorch can prompt users to set this parameter in \textit{Upsample} to keep the program running properly. 
% In addition to this, \tool~detects that Jittor has issues with calling \textit{jt.softplus}, which has not yet been implemented, and misspelling of API names.

\begin{figure}[h]
  \centering
  \includegraphics[width=\linewidth]{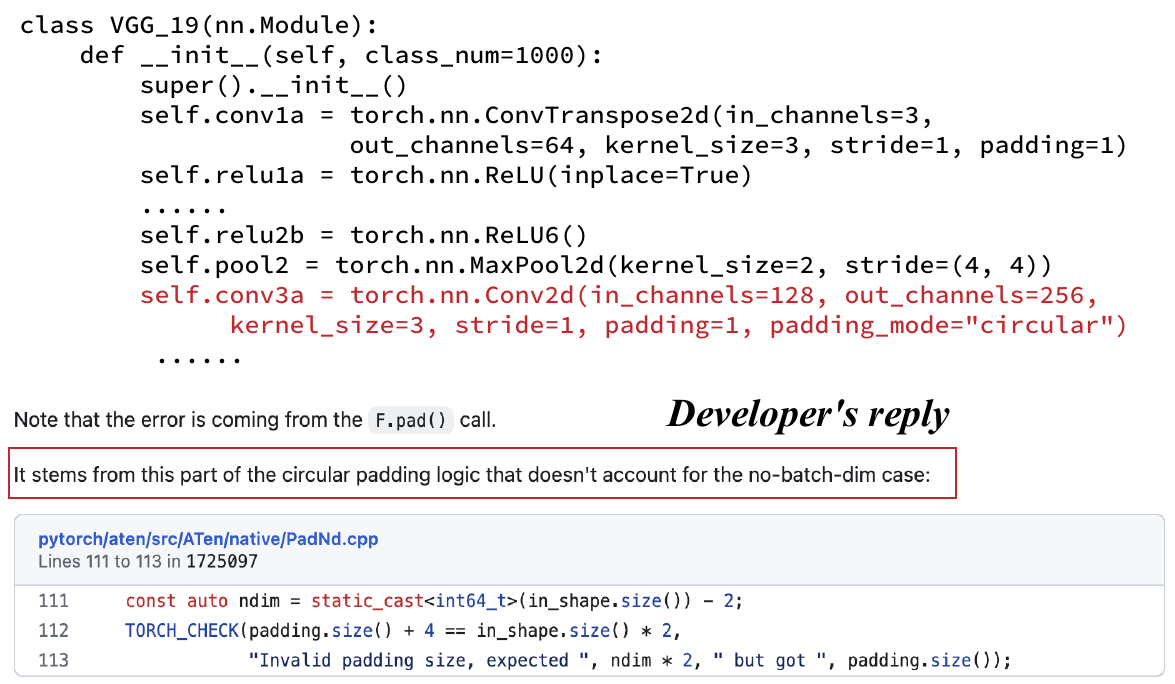}
  % \captionsetup{skip=0.3em}
  \caption{Implementation bug in torch.nn.Conv2d}
  \label{fig:impbug-torch}
\end{figure}

As shown from the code snippet using PyTorch in Figure \ref{fig:impbug-torch}, when using the random generation operator on \textit{torch.nn.Conv2d}, \tool~detects that 3-dim input cannot work properly when \textit{padding\_mode} is set to \textit{circular}. 
The developer confirms that this bug is due to the logic of the circular padding not taking into account the case of no-batch-dim. 
To date, this bug has been fixed. 
While testing with the PointNet model, we find that \textit{torch.nn.Conv1d} has the same bug when feeding the 2-dim input. 
% FreeFuzz does not detect this bug, and we analyze that it may be due to the strong code derivation capability of \tool, which allows APIs to be tested with many more combinations of parameter values than FreeFuzz.
% DeepREL cannot detect this bug because relational APIs are likely to be using the same primitive computation operators. 
% When comparing the outputs of Conv1d and Conv2d, DeepREL cannot detect the bug that they both have. 
The reasons why DeepREL cannot detect this bug may be as follows: (1) the inputs of tested APIs are generated randomly rather than from flow dependency, so the input dimension and characteristics of the datasets in real application scenarios are not considered, resulting in the triggering conditions of this bug not being met. (2) Relational APIs are likely to use the same primitive computation operators, so when they have the same potential bug, that bug cannot be detected. 

\textbf{Performance Bug.} \tool~employs dynamic packetization to import newly generated code for implementing DL models in the same testing process, indicating that multiple DL models are constructed and trained one after the other. 
During this process, TensorFlow throws the error \textit{ResourceExhaustedError: failed to allocate memory}.
After analysis, we find that TensorFlow does not clean up the graphics memory and other system resources occupied by building and training the DL model when a single model is trained, resulting in the graphics memory growing and causing the \textit{OOM} issue. 
We have tried several performance optimizations, such as using \textit{tf.keras.backend.clear\_session()}, but none of them worked. 
PyTorch and Jittor have no similar issues. 
We have submitted this performance bug, and the developers are trying to fix it.
Since API-level fuzzing methods do not involve training, it is difficult to identify the performance bug that requires training the model to expose.

\subsection{Discussion}

% For the model-level fuzzing methods, new models are generated by removing and adding layers in the seed models, and they require that the input shape and the output shape of the selected layer be the same.
Currently, there are two main types of test case generation methods in model-level DL library testing methods: (1) removing and adding layers in seed models and (2) generating models based on highly abstract model structures. 
The former requires the input shape and the output shape of the selected layer to be consistent, which is limited by the structures of the seed model~\cite{wang2020deep}.
The latter only supports a certain range of APIs to ensure the generated models can work properly. 
Therefore, they both are limited by the predefined configurations. 
% For model-level fuzzing methods, they generate new models by removing and adding layers in the seed models or based on highly abstract model structures. 
% In either case, they are limited by predefined configurations (e.g., the structures of the seed model~\cite{wang2020deep} and the scope of the adapted APIs~\cite{gu2022muffin}). 
% Therefore, the type and number of APIs and the lines of code covered by the generated tests during testing are limited by the seed models.
Unlike model-level fuzzing testing methods that mutate DL models, \tool~takes the code as the mutation object to generate complete code files that implement models of various structures. 
Template-based code assembly has more maneuverable space than directly mutating a trained model and provides more guidance information that can ensure the quality of tests than using randomly generated abstract structures. 
The combination of code assembly and well-designed mutation operators can effectively improve testing efficiency and achieve high testing adequacy.

\begin{figure}[h]
  \centering
  \includegraphics[width=\linewidth]{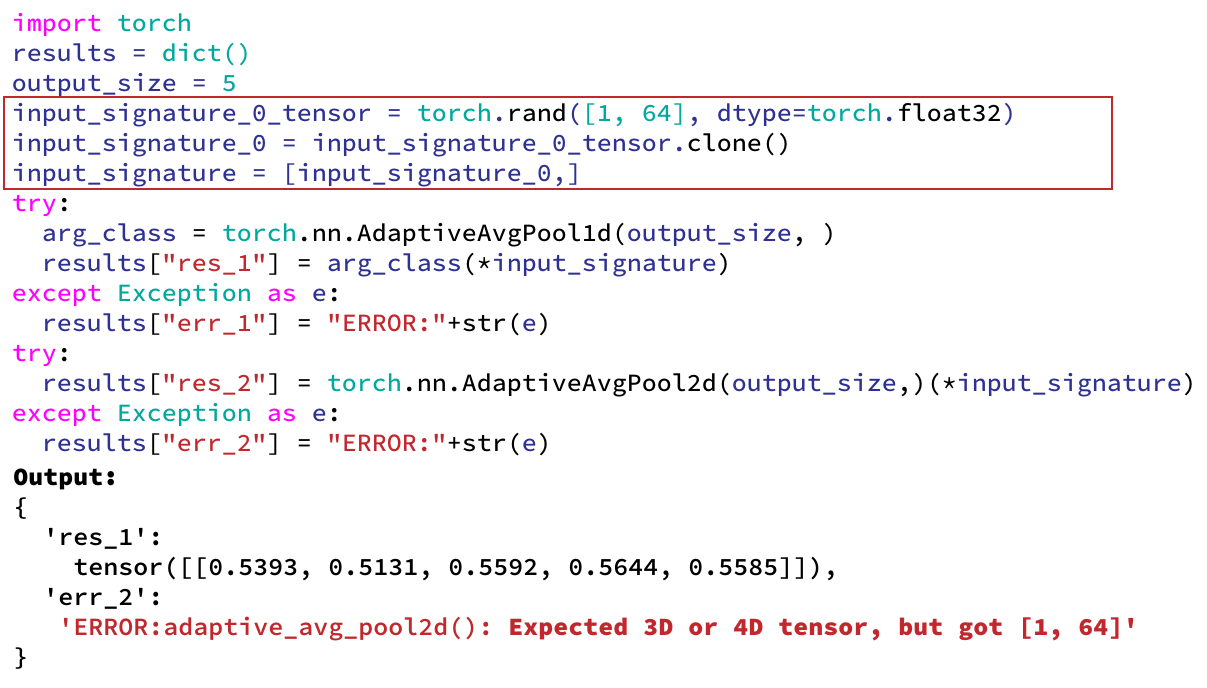}
  % \captionsetup{skip=0.3em}
  \caption{The example of false positives output by DeepREL}
  \label{fig:deeprel-test}
\end{figure}

In contrast to API-level fuzzing methods, the tests generated by \tool~are not just a few lines of code but a complete code file for implementing the DL model that is rich in semantics and supports testing with inputs in real application scenarios. 
Moreover, we find that API-level fuzzing methods still need to be improved in terms of test oracle construction and precision. 
Even though we use the optimal parameter values provided by the authors, DeepREL still outputs many false positives, which are caused by incorrectly matched relational APIs, incorrect parameter values, mismatched input dimensions, etc. 
As shown in Figure~\ref{fig:deeprel-test}, many of the false positives output by DeepREL are due to the wrong shape of input data generated for testing the API. 
The example given in Figure~\ref{fig:deeprel-test} is where DeepREL incorrectly assumes that the input shapes for \textit{torch.nn.AdaptiveAvgPool1D} and \textit{torch.nn.AdaptiveAvgPool2D} should be the same.
Therefore, when checking the outputs of DeepREL, it is imperative to allocate a substantial amount of time to verify the validity of the relation in the API pair, confirm that there is no issue with the syntax and semantics of the generated test cases, identify which API in the pair has a potential bug, confirm which parameter value triggers the bug, and so on. 
This problem also exists in the application of Muffin.
Because the test cases generated by Muffin are independent of each other and have low readability, too little information can be provided for bug localization, and only a few lines of the error information in the log can be used for manual reasoning and review.

% Therefore, when checking the outputs of DeepREL, we need to spend a lot of time confirming whether the relation between the API pair is valid, which API in the pair has a potential bug, which parameter value induces the bug, and so on. 
% The benefit of this design is that \tool~can trigger more bugs that a single line of code cannot trigger by using context-based data flow.
% Also, the performance of the API-fuzzing method is affected by the parameter settings. 
% Even though we use the same parameter setting as in the paper on FreeFuzz, FreeFuzz still cannot trigger some bugs in boundary value checking.
% \tool~can make full use of the information provided by the seed, and even if the number of mutation times of the code block is set to 5, it still has excellent fuzzing performance over FreeFuzz. 
% To ensure the validity of generated tests, \tool~adopts the idea of disassembling the seed test as a basis for assembling new tests.
% The outputs of the disassembling step consist of a template and code blocks.
% The template is used to guide the generation of tests to ensure semantic-syntactic correctness.
% Code blocks are accessories to be subsequently assembled and can be mutated to cover more APIs and lines of code.
Although we cannot guarantee that the generated code files are 100\% syntactically and semantically correct, the precision of \tool~is 68.7\% after manual review.
After the experimental analysis and performance comparison, we have reasons to believe that the validity of tests generated by~\tool~is acceptable. 
In this study, we focus on APIs of DL libraries related to implementing DL models. 
Implementing the DL model in a DL project involves crucial steps such as selecting appropriate layers, activation functions, parameter settings, etc.
Therefore, from the user requirements perspective, focusing on testing these core APIs ensures that the critical functionality used to support the DL model implementation is effectively validated to meet the actual needs of the majority of users.
% The mutation objects of \tool~are code, so it is highly extensible.
\tool~uses code files as seed tests, making it highly extensible. 
The API mutation operators of \tool~are implemented based on information provided by the document, so as long as there is a document that can give the definition of the API, the parameter list of the API, the definitions, and the legal value ranges of API parameters, we can quickly implement the mutation methods suitable for various APIs. 
Due to the flexibility and extensibility of \tool, we believe \tool~can be easily adapted to seed code files for other application scenarios of DL libraries (e.g., data preprocessing and augmentation, model deployment and optimization, etc.) to cover more APIs in deep learning libraries.
% Therefore, \tool~can easily be adapted to seed code for other application purposes and cover more APIs of DL libraries.
This is also our future work.

\subsection{Threats to Validity}
\subsubsection{The threats to internal validity}
The threats to internal validity of \tool~mainly come from the test generation and ground-truth construction. 
We obtain useful information from official documents to implement the code mutation operators.
However, in the official documentation, the legal value range of some parameters is very vague. 
For example, for some parameters of integer type, the documentation does not give a clear legal minimum or maximum value.
This causes \tool~to occasionally generate legitimate but illogical parameter values, resulting in mutated code blocks that cannot work properly when placed in context.
In the phase of ground-truth construction, each error report is assigned to three authors and carefully reviewed to determine whether it contains a bug, the type of bug it contains, or the type of false positive. When disagreements arise, the authors discuss them together to resolve them.

\subsubsection{The threats to external validity}
The threats to external validity of \tool~mainly come from the tested DL libraries and used seed tests.
We select three commonly-used DL libraries, i.e., TensorFlow, PyTorch, and Jittor, as subjects.
In contrast to Muffin, \tool~is not limited by the need for the front-end of the subject to be Keras, and PyTorch and Jittor are also supported. 
Also, the fuzzing method we proposed is general for other types of code files.
As long as the mutation operators are re-implemented based on the corresponding official documentation and then re-implement the code disassembly and assembly methods according to the code style, \tool~can be used to test any other DL libraries and any code files for other purposes.
% In the experimental phase, the seed tests we selected can build 9 DL models of different types and structures.

\section{Related Work}
% This section introduces the research work related to \tool, which is divided into the testing methods of the DL library and the testing methods of the DL model.

\subsection{Deep Learning Library Testing}
% \textbf{Deep Learning Library Testing.} 
Currently, testing methods for DL libraries can be mainly classified into two categories: model-level testing methods and API-level testing methods. 
They primarily employ fuzzing testing techniques, with the former utilizing DL models as seed data and the latter using code that calls the APIs as seed data.
LEMON~\cite{wang2020deep}, AUDEE~\cite{guo2020audee}, and Muffin~\cite{gu2022muffin} represent the state-of-the-art model-level testing methods.
LEMON generates new models based on the seed model by adding/removing layers of the seed and changing the value of the weight tensor.
AUDEE implements three distinct mutation strategies to explore combinations of model structures, parameters, weights, and inputs to generate diverse tests.
They both employ differential testing methods to construct test oracles, such as comparing the outputs of the same structural model on different libraries.
% For example, bugs can be detected by checking whether models constructed with different DL Libraries have the same output after feeding the same input. 
This test oracle imposes requirements on the accuracy of newly generated models and can also inversely constrain test input generation. 
% Muffin first generates the DL model structure information and then generates information about each layer based on predefined supported APIs, so the maximum number of APIs it can cover is the number of supported APIs. 
To ensure that the generated models work, Muffin selects layers that adapt the randomly generated abstract structure from the predefined supported APIs to generate the model. 
Thus, the diversity of test inputs is limited by the number of supported APIs. 
% Therefore, there are limitations in the process of model generation and mutation, and the final mutation results cannot cover a wider range of APIs.
FreeFuzz~\cite{10.1145/3510003.3510041} and DocTer~\cite{10.1145/3533767.3534220} are API-level fuzz testing methods that extract information from official documents and open-source code. 
DeepREL~\cite{deng2022fuzzing} further improves FreeFuzz by inferring relational APIs automatically and fuzzing the DL libraries based on their relations.
However, they cannot fully evaluate how well the DL Library handles illegal inputs because of the supporting data source, and they still need to improve precision. 
In addition, there are some works that test specific components of deep learning libraries. 
For example, Yang et al. propose the testing tool $\nabla$Fuzz for automatic differentiation components~\cite{yang2023fuzzing}, and Liu et al. propose the testing tool NNsmith for deep learning compilers~\cite{liu2023nnsmith}.

\subsection{Deep Learning Model Testing}
% \noindent\textbf{Deep Learning Model Testing.} 
DL model testing is mainly concerned with whether the output of the model is correct. 
Currently, many studies have proposed effective methods for detecting erroneous outputs of DL models in various application domains, e.g., natural language processing~\cite{10.1109/ASE51524.2021.9678715,chen2021testing}, speech recognition~\cite{ji2022asrtest, yuen2023asdf}, image recognition~\cite{tian2020testing, wang2020test}, object detection~\cite{wang2020metamorphic, guo2022lirtest}, and autonomous driving~\cite{li2022comopt, nguyen2021salvo}.
In addition, some studies focus on measuring test adequacy~\cite{pei2017deepxplore, ma2018deepgauge,sun2019structural,ma2019deepct} and prioritizing test cases for the DL model~\cite{kim2019guiding,feng2020deepgini}, aiming to improve testing efficiency further.
% In addition, some studies focus on the measurement of test adequacy~\cite{pei2017deepxplore, ma2018deepgauge,sun2019structural,ma2019deepct} and prioritizing test cases for the DL model~\cite{kim2019guiding,feng2020deepgini}, with the aim of further improving the testing efficiency.
% Unlike them, we test DL libraries by simulating the process of acquiring DL models (building, training, and evaluating). 
Unlike them, we test the fundamental component that significantly influences the quality of DL systems rather than focusing on their application level.
Therefor, DL models are intermediates in the execution of \tool~rather than the direct subject under test.

% \vspace{-1.0em}
\section{Conclusion}
In this paper, we propose \tool, a novel fuzzing testing method for DL libraries, which can generate new code files for implementing DL models with different structures via assembling code. 
% \tool~aims to simulate the process of constructing, training, and evaluating DL models, which are the most common real-world user scenarios of DL libraries.
% We implement this method in the testing tool~\tool.
The code files generated by \tool~are related to the process of constructing, training and evaluating DL models, which are the most common real-world user scenarios of DL libraries.
\tool~first disassembles the seed into the initial template and multiple code blocks and then uses code block mutation operators to generate more code blocks that can be adapted to the template. 
With the generation guidance provided by the template, \tool~can eventually produce a derivation tree of code files.
We construct the test oracle based on the derivation relation in this tree.
By executing the newly generated code files, \tool~can detect and locate bugs.
% If an anomaly such as a crash occurs while running the generated code file, it indicates that there may be a bug. 
During the experiment, \tool~detects 64 new bugs of four types in the latest release versions of three DL libraries (e.g., TensorFlow, PyTorch, and Jittor), where 51 bugs have been confirmed, and 13 bugs have been fixed by developers.
The experimental results demonstrate that \tool~has excellent capabilities for generating high-quality tests and detecting different types of bugs in multiple DL libraries.

% \vfill
\bibliographystyle{IEEEtran}
\balance
\bibliography{ref.bib}
\end{document}